\let\orilabel\label 
\documentclass[reprint,amsmath,amssymb,aps,pre,showkeys,superscriptaddress,frontmatterverbose]{revtex4-2}
\let\label\orilabel 

\usepackage{graphicx} 
\usepackage{dcolumn} 
\usepackage{bm} 

\usepackage[mathlines]{lineno}

\usepackage{multirow}
\usepackage{array}
\usepackage{makecell}

\usepackage{enumitem}

\usepackage{amsfonts}
\usepackage{bbm}   

\newcommand{\tc}{\overline{c}}               
\newcommand{\nhat}{\widehat{\mathbf{n}}}     
\newcommand{\tf}[1]{\widetilde{#1}}          
\newcommand{\bb}[1]{\mathbb{#1}}             
\newcommand{\bbm}[1]{\mathbbm{#1}}           

\newcommand{\avg}[1]{\langle #1 \rangle}      
\newcommand{\avgt}[1]{\langle \widetilde{#1} \rangle}  

\newcommand{\Dbb}{\mathbb{D}}                          
\newcommand{\Deff}{\avgt{\Dbb}_{eff}}                  
\newcommand{\Dnd}{\avgt{\Dbb}}                         

\newcommand{\Weff}{\avgt{\mathbf{W}}_{eff}}            
\newcommand{\Nnd}{\avgt{\mathbf{N}}}                   

\newcommand{\heff}{\avgt{h}_{eff}}                     
\newcommand{\Shnd}{\avgt{Sh}}                          

\newcommand{\Reff}{\avgt{\mathbf{R}}_{eff}}            
\newcommand{\Rnd}{\avgt{\mathbf{R}}}                   

\newcommand{\Nx}{\avgt{N}_x}
\newcommand{\Rx}{\avgt{R}_x}
\newcommand{\Dxx}{\avgt{\Dbb}_{xx}}
\newcommand{\Dyy}{\avgt{\Dbb}_{yy}}

\newcommand{\tortu}{\avgt{\bbm{t}}}     
\newcommand{\hdisp}{\avgt{\bbm{d}}}     

\newcommand{\Xg}{\mathbf{X}}     
\newcommand{\xl}{\mathbf{x}}     
\newcommand{\uu}{\mathbf{u}}     
\newcommand{\up}{\mathbf{u}'}    
\newcommand{\jj}{\mathbf{j}}     
\newcommand{\bcv}{\mathbf{b}}    

\newcommand{\Da}{Da}                          
\newcommand{\Pe}{Pe}                          
\newcommand{\wstar}{\omega^{*}}               

\newcommand{\lP}{l_P}
\newcommand{\lRVE}{l_{RVE}}
\newcommand{\Lsys}{L_{sys}}
\newcommand{\Vf}{V_f}
\newcommand{\Vs}{V_s}
\newcommand{\VRVE}{V_{RVE}}
\newcommand{\sA}{s_A}
\newcommand{\eps}{\varepsilon}

\newcommand{\cs}{c_{i,s}}         
\newcommand{\cseq}{c_{i,s,eq}}    
\newcommand{\ki}{k_i}             

\newcommand{\Ny}{\avgt{N}_y}
\newcommand{\Ryc}{\avgt{R}_y}
\newcommand{\Dxy}{\avgt{\Dbb}_{xy}}
\newcommand{\Dyx}{\avgt{\Dbb}_{yx}}

\newcommand{\iu}{j}               

\usepackage[hidelinks]{hyperref} 
\usepackage{pdfpages}
\makeatletter
\AtBeginDocument{\let\LS@rot\@undefined}
\makeatother

\begin{document}

    \preprint{APS/123-QED}
    \title{An Omni-Temporal Framework for Hydrodynamic Dispersion and Reaction in Porous Media} 

\author{Md Abdul Hamid}
\email{mahamid.me.buet@gmail.com}
\affiliation{Department of Mechanical Science and Engineering, University of Illinois Urbana-Champaign, Urbana, Illinois 61801, USA}
\affiliation{Department of Mechanical Engineering, Clemson University, Clemson, South Carolina 29634, USA}

\author{Kyle C. Smith}
\affiliation{Department of Mechanical Science and Engineering, University of Illinois Urbana-Champaign, Urbana, Illinois 61801, USA}
\affiliation{Department of Materials Science and Engineering, University of Illinois Urbana-Champaign, Urbana, Illinois 61801, USA}
\affiliation{Computational Science and Engineering Program, University of Illinois Urbana-Champaign, Urbana, Illinois 61801, USA}

\date{\today}
    \begin{abstract}
Many applications involve porous-media transport driven on time scales comparable to the pore-scale diffusion time, for which classical pseudo-steady dispersion models based on time-invariant transport coefficients may become inaccurate. While frequency-dependent dispersion has been studied in non-reactive porous media, a corresponding framework that simultaneously accounts for advection, diffusion, and heterogeneous interfacial reaction remains lacking. We address this gap by introducing an omni-temporal framework based on volume averaging of the frequency-domain pore-scale advection-diffusion equation subject to a Robin boundary condition spanning the full range of interfacial reactivity. The resulting upscaled formulation yields four frequency-dependent transfer functions: a dispersion tensor, an advection-suppression vector, a spectral Sherwood number, and a reactivity-bias vector, with the reactivity-bias vector introduced here for the first time. Analytical solutions for fully developed flow between parallel plates and through circular tubes reveal distinct kinetics- and diffusion-limited regimes, including capacitive and semi-infinite Warburg-type frequency responses. Comparisons of transient breakthrough curves with direct numerical simulations demonstrate substantial improvement over conventional pseudo-steady dispersion, particularly for short-duration pulses. Application to a periodic array of square rods under cross flow further demonstrates the framework's applicability beyond canonical geometries.
\end{abstract}
    \maketitle

    \section{\label{sec:Sec_01}Introduction}

\begin{figure}[t!]
    \centering
    \includegraphics[width=0.9\linewidth]{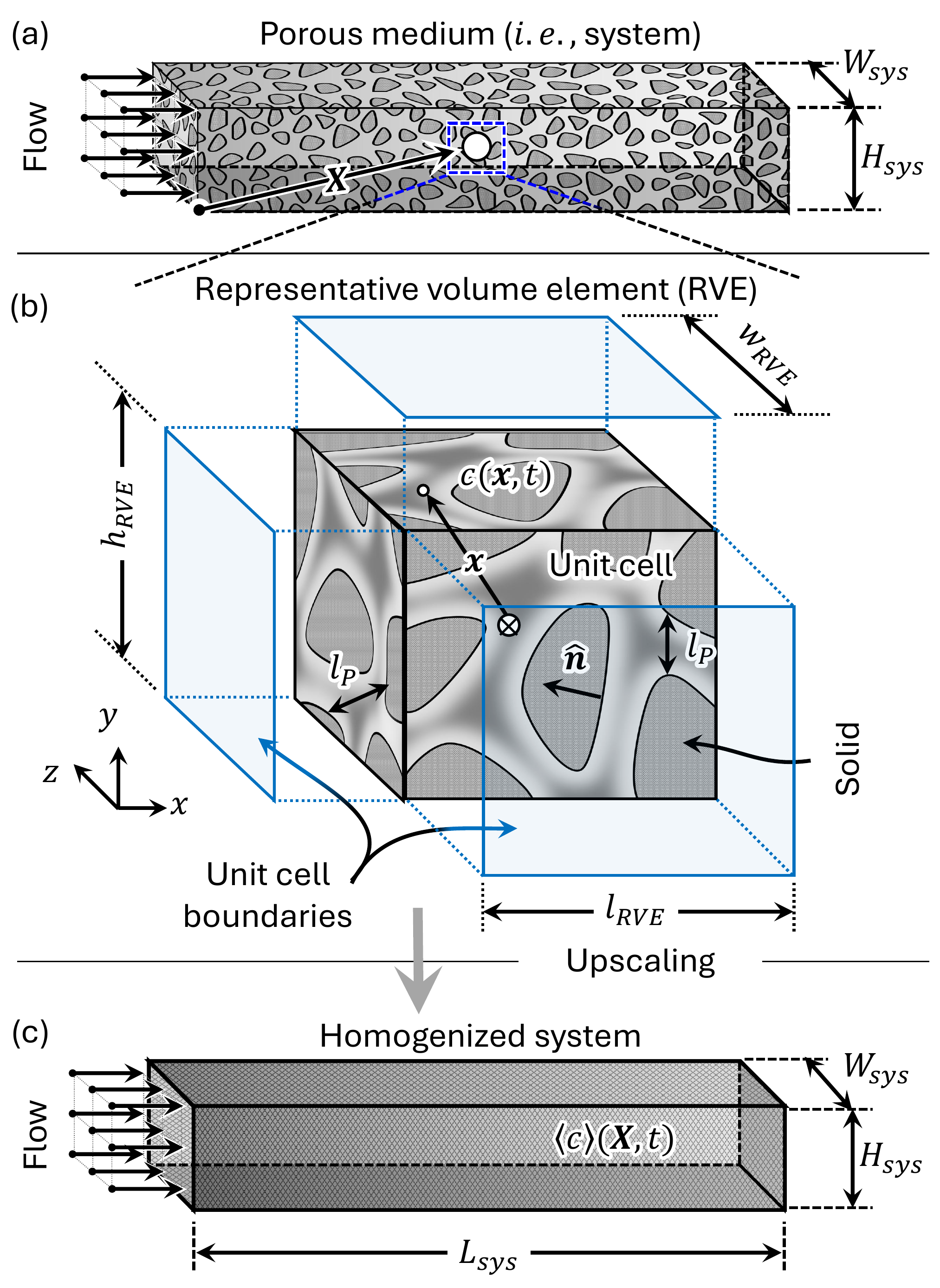}
    \caption{Schematic illustration of porous-media upscaling using an RVE. (a) Porous medium (system) of characteristic dimensions $\Lsys\times H_{sys}\times W_{sys}$, where the global position vector $\Xg$ locates an RVE within the system. (b) Periodic RVE of dimensions $\lRVE\times h_{RVE}\times w_{RVE}$ containing solid (gray) and fluid (pore) regions, where the local position vector $\xl$ specifies the pore-scale position associated with the concentration field $c(\xl,t)$. The characteristic pore length $\lP$ and outward unit normal $\nhat$ are also shown. (c) Homogenized system described by the local volume-averaged concentration field $\avg{c}(\Xg,t)$. Although the system and RVE are illustrated as prismatic, the upscaling framework is not restricted to these geometries.}
    \label{fig:1}
\end{figure}

\begin{figure}[t!]
    \centering
    \includegraphics[width=0.85\linewidth]{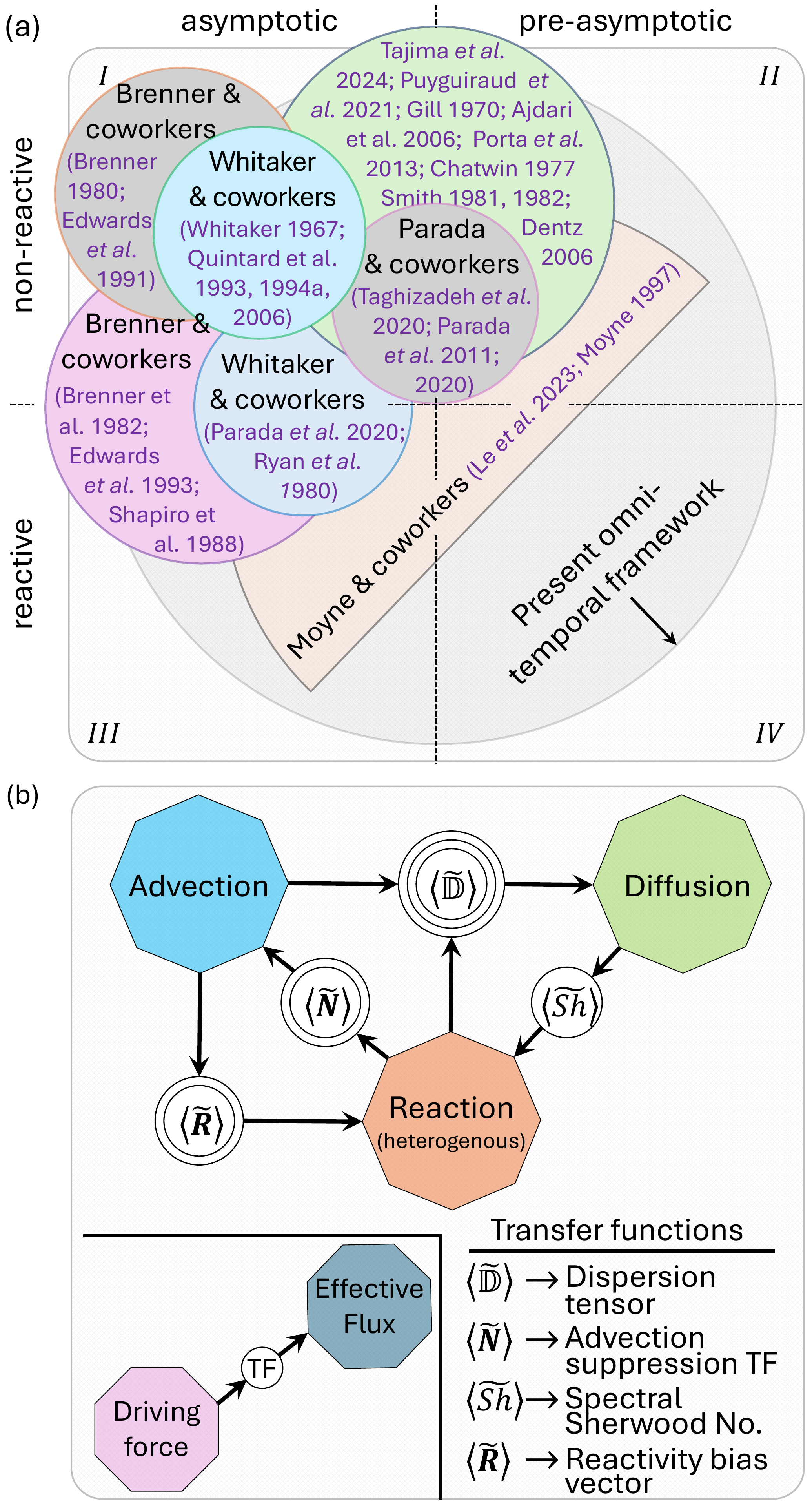}
    \caption{(a) Schematic classification of representative dispersion frameworks according to interfacial reactivity and temporal regime. The upper and lower halves correspond to non-reactive and reactive solid/liquid interfaces, respectively, while the left and right halves represent long-time (asymptotic) and short-time (transient or pre-asymptotic) transport, respectively. Representative studies are placed within quadrants I--IV according to their underlying assumptions. Full circles denote studies that account for both advection and diffusion (hydrodynamic dispersion); marker size carries no physical significance. (b) Schematic mapping between driving forces and effective fluxes through transfer functions (TFs), illustrated generically in the lower-left tile. The present framework describes coupled advection, diffusion, and heterogeneous interfacial reaction through four TFs. Each TF relates a driving force to its corresponding effective flux; for example, $\Nnd$ describes the reaction-induced modification of effective advection, while $\Shnd$ characterizes interfacial mass transfer. Scalar-, vector-, and tensor-valued TFs are represented by single, double, and triple circles, respectively.}
    \label{fig:2}
\end{figure}

Solute transport through porous media governs a broad range of electrochemical, geological, and biological processes, including chromatographic separation \cite{Giddings1965}, flow batteries \cite{Wlodarczyk2024,Lisboa2017,Nemani2017}, deionization cells \cite{Do2023,Juchen2023,Lee2019,Stroman2017,Rahman2025}, catalytic converters \cite{Stary2006,Hayes2012}, coastal aquifers \cite{Wang2024,Alfarrah2018,Badaruddin2015,Prusty2020}, enhanced oil recovery \cite{Blackwell1959,GamalRezk2022}, $\mathrm{CO_2}$ sequestration \cite{Zevenhoven,Kalam}, radioactive waste repositories \cite{Rashwan2022,Amaziane2012}, pulmonary airflow \cite{DeGroot2011,Koulich1999,Kuwahara2009,AvilesRojas2022}, and drug delivery systems \cite{Nicholson2001,Khaled,Siepmann}. When transport conditions vary much more slowly than the pore-scale diffusion time, the pore-scale concentration field has sufficient time to adjust to macroscale forcing. The effective transport coefficients then approach time-independent values, defining the pseudo-steady, or quasi-steady, limit associated with asymptotic dispersion. In many applications, however, transport conditions vary on time scales comparable to, or shorter than, the pore-scale diffusion time. Examples include cyclic charge and discharge in flow batteries, pulsed or intermittent injection in groundwater remediation and enhanced oil recovery, and oscillatory ventilation in the lungs. Under these conditions, the pore-scale concentration field cannot fully adjust to the changing macroscale forcing, and the effective transport coefficients become time- or frequency-dependent. This corresponds to the pre-asymptotic, or transient, dispersion regime. Applying pseudo-steady transport coefficients in this regime can substantially mispredict breakthrough times, peak concentrations, and interfacial reaction rates, with direct consequences for the prediction, design, and control of transient porous-media systems. A transient upscaling framework is therefore needed when the imposed transport time scale becomes comparable to the pore-scale diffusion time.

The large separation between pore and system length scales motivates the use of upscaling methods. These methods replace the detailed pore-scale concentration field with a local volume-averaged field governed by effective transport coefficients. The resulting macroscale transport is commonly described by an advection-dispersion-reaction equation (ADRE) defined over a homogenized porous medium. Figure~\ref{fig:1} illustrates this concept. The original porous medium in Fig.~\ref{fig:1}(a) is represented by the homogenized system in Fig.~\ref{fig:1}(c), while the effective transport coefficients are determined from pore-scale conservation equations solved over the RVE shown in Fig.~\ref{fig:1}(b).

A central challenge with upscaling procedures is to represent the effects of pore-scale concentration heterogeneity within the homogenized ADRE. Taylor~\cite{Taylor1953} established the foundation for hydrodynamic dispersion by showing that the long-time, cross-sectionally averaged spreading of a solute in Poiseuille flow through a tube can be represented by an enhanced axial diffusivity arising from the interaction between the nonuniform velocity field and transverse molecular diffusion. Brenner and co-workers subsequently developed a probabilistic framework based on Green's-function solutions of the transient pore-scale advection-diffusion equation subjected to impulse forcing \cite{Brenner1980,Edwards1991}. Positional moments of the resulting concentration field were then used to determine effective, or phenomenological, transport coefficients. In parallel, Whitaker and co-workers \cite{Whitaker1967,Quintard1994a,Quintard1993,Quintard2006} developed a volume-averaging approach in which the pore-scale conservation equations are averaged directly to obtain the macroscale governing equations, effective transport coefficients, and associated closure problems. Volume averaging has since been applied broadly to mass and heat transport and to the analysis and design of porous microstructures \cite{Kaviany2001,Kaviany1995,Calmidi2000,Ozgumus2013}. Detailed treatments of the probabilistic and volume-averaging approaches are given by Brenner and Edwards~\cite{BrennerEdwards1993} and Whitaker~\cite{Whitaker1999}, respectively.

Although these approaches can effectively predict transport for a prescribed porous microstructure, classical formulations are generally associated with the asymptotic regime, where the characteristic transport time scale is substantially longer than the pore-scale diffusion time, $\tau_d\equiv l_c^2/D$. Here, $l_c$ and $D$ denote the characteristic pore length and molecular diffusivity, respectively. Several studies have extended dispersion models into the pre-asymptotic regime, but a general formulation that simultaneously accounts for advection, diffusion, and heterogeneous interfacial reaction across both temporal regimes remains lacking.

Figure~\ref{fig:2}(a) summarizes this gap by classifying representative dispersion frameworks according to temporal regime and interfacial reactivity. Most previous studies are concentrated in the non-reactive, asymptotic limit (quadrant I). Studies spanning quadrants I and II \cite{ValdesParada2011,Taghizadeh2020,Porta2013,Tajima2024,Puyguiraud2021,Gill1970,Chatwin1977,Ajdari2005,Smith1981,Smith1982,Dentz2006} extend dispersion into the pre-asymptotic regime but remain restricted to non-reactive systems governed by coupled advection and diffusion. Among these, Smith~\cite{Smith1981,Smith1982} and Dentz and Tartakovsky~\cite{Dentz2006} use delay-diffusion approaches in which the deviation concentration depends on the history of the macroscale concentration gradient through an assumed or truncated memory representation. These approaches capture memory associated with coupled advection and diffusion but do not include heterogeneous interfacial reaction. Conversely, studies spanning quadrants I and III \cite{Shapiro1988,BrennerAdler1982,Edwards1993,ValdesParada2020,Ryan1980} incorporate interfacial reaction but remain restricted to asymptotic transport. Approaches spanning all four quadrants have considered asymptotic and pre-asymptotic reactive transport, but for diffusion-only systems without advection \cite{Moyne1997,Le2023}. Thus, a gap remains for an upscaled framework that simultaneously describes advection, diffusion, and heterogeneous interfacial reaction across both asymptotic and pre-asymptotic regimes.

The present work addresses this gap through an {\em omni-temporal} dispersion framework based on volume averaging of the Fourier-transformed pore-scale governing equations. We use the term {\em omni-temporal} because the framework describes transport across time scales within a single formulation, from the fast, pre-asymptotic (transient) regime to the slow, long-time asymptotic (pseudo-steady) limit. A single Robin-type boundary condition spans the full range of interfacial reactivity, characterized by the Damk\"ohler number $\Da\equiv\text{reaction rate}/\text{diffusion rate}$, from the non-reactive limit ($\Da=0$) to the perfectly reactive limit ($\Da\to\infty$). The framework therefore simultaneously accounts for advection, diffusion, and heterogeneous interfacial reaction across the four transport regimes illustrated in Fig.~\ref{fig:2}(a).

The proposed framework yields four frequency-dependent upscaled transport coefficients: a dispersion tensor $\Dnd$, an advection-suppression transfer function $\Nnd$, a spectral Sherwood number $\Shnd$, and a reactivity-bias vector $\Rnd$. The reactivity-bias vector is introduced here for the first time to quantify how flow redistributes solute toward or away from reactive interfaces prior to reaction, thereby characterizing the coupling between advection and interfacial reactivity that is not explicitly represented in existing dispersion frameworks. The four coefficients can be interpreted as transfer functions (TFs) relating driving forces to their corresponding effective fluxes, as illustrated in Fig.~\ref{fig:2}(b). The number and form of the required TFs depend on the transport processes present. When reaction is absent, for example, the reactive couplings vanish and the formulation reduces to the dispersion tensor $\Dnd$, which describes the combined influence of pore-scale advection and diffusion on macroscale dispersive transport.

The frequency-domain formulation also provides a direct representation of the memory inherent to transient transport. Previous pre-asymptotic upscaling studies have represented time-dependent transport coefficients \cite{ValdesParada2011,DavitQuintard2012,ValdesParada2012} and memory through temporal convolution kernels \cite{Cushman2000,Chastanet2008} or explicit delay-diffusion formulations \cite{Smith1981,Smith1982}. In the present framework, memory arises naturally from the frequency dependence of the transfer functions: after inverse Fourier transformation, their products with the corresponding driving forces become temporal convolutions. The memory kernels are therefore determined directly from the pore-scale closure problems rather than prescribed independently. Related frequency-based formulations were previously developed by us for the perfectly reactive limit under steady flow in redox-flow batteries \cite{Hamid2020,Hamid2023}. In those studies, pseudo-steady and time-domain responses agreed well with transient electrochemical experiments, despite neglecting hydrodynamic dispersion by assuming a zero macroscale concentration gradient. Following White and co-workers \cite{Hu2012a,Hu2012b}, the time-domain responses were reconstructed using state-space approximations of the numerically determined $\Shnd$ and $\Nnd$. The present work generalizes this frequency-based approach by incorporating nonzero macroscale concentration gradients and the full range of interfacial reactivity, thereby simultaneously accounting for transient advection, diffusion, and heterogeneous interfacial reaction in porous media.

The principal findings are as follows. First, the omni-temporal framework unifies dispersion, reaction-induced advection suppression, interfacial mass transfer, and flow-induced reactivity bias through four frequency-dependent transfer functions that span the full range of interfacial reactivity and temporal response. Second, the results show that the tortuosity contribution to effective transport is process-dependent: the tortuosity tensor vanishes in the perfectly reactive limit because reactive interfaces act as solute sources or sinks rather than solely as barriers to diffusion. This result provides a connection between geometric and process-dependent interpretations of tortuosity \cite{Ghanbarian2013,Holzer2023}. Third, analytical solutions for Poiseuille flow between parallel plates and through circular tubes reveal distinct kinetics- and diffusion-limited regimes. At sufficiently small Damk\"ohler number, the reactive coefficients exhibit their kinetics-limited asymptotic behavior, whereas the diffusion-limited regime exhibits characteristic Warburg-type frequency scaling in $\Nnd$, $\Shnd$, and $\Rnd$, analogous to semi-infinite diffusion in electrochemical impedance \cite{Huang2018}. The dispersion tensor exhibits a distinct high-frequency response, approaching the purely diffusive limit while its phase varies non-monotonically before returning toward zero. Fourth, comparison with direct numerical simulations of transient solute pulses shows that the frequency-dependent formulation substantially improves predictions of pulse propagation and axial spreading relative to conventional pseudo-steady dispersion in the pre-asymptotic regime. Finally, application to a periodic array of square rods under cross flow demonstrates that the same transfer-function structure extends beyond the canonical parallel-plate and circular-tube geometries.

The remainder of this paper is organized as follows. The pore-scale mass conservation equations are formulated in the time domain in Sec.~\ref{sec:Sec_02}. Section~\ref{sec:Sec_03} transforms these equations into the frequency domain and applies local volume averaging to derive the closure problems, frequency-dependent transport coefficients, and a homogenized ADRE. Section~\ref{sec:Sec_04} presents analytical expressions for fully developed flow between parallel plates and through circular tubes across the full range of interfacial reactivity. Section~\ref{sec:Sec_05} compares the upscaled predictions with direct numerical simulations, and Sec.~\ref{sec:Sec_06} extends the framework to periodic arrays of square rods under cross flow.
    
    \section{\label{sec:Sec_02}Pore-Scale Transport in the Time Domain}

We consider convective solute transport through a porous medium. The global position vector $\Xg$ specifies the location of an RVE within the porous medium [Fig.~\ref{fig:1}(a)], while the pore-scale position vector $\xl$ specifies the local position within the RVE [Fig.~\ref{fig:1}(b)]. The characteristic RVE length satisfies $\lRVE\gg\lP$, where $\lP$ is the characteristic pore length, providing separation between the pore and RVE length scales. The total RVE volume $\VRVE$ consists of solid and fluid volumes, $\Vs$ and $\Vf$, such that $\VRVE=\Vs+\Vf$. The porous medium is assumed rigid and macroscopically homogeneous, with negligible spatial variation in porosity, $\eps\equiv\Vf/\VRVE$. Heterogeneous reactions occur at the solid/liquid interface within the RVE [Fig.~\ref{fig:1}(b)], while homogeneous reactions are neglected. The interfacial geometry is assumed to remain unchanged by the heterogeneous reaction.

The pore-scale mass conservation equation for species $i$ is
\begin{equation}
    \frac{\partial c_i}{\partial t} + \nabla\cdot\mathbf{j}_i = 0. \label{eq:1}
\end{equation}
Here, $c_i(\xl,t)$ is the pore-scale concentration and $\jj_i$ is the species flux. In general, the flux contains advective, diffusive, and electromigrative contributions, such that $\jj_i=\jj_i^a+\jj_i^d+\jj_i^m$. The advective flux is $\jj_i^a=\uu c_i$, where the local velocity field $\uu(\xl)$ is generated by pressure- or gravity-driven flow through fully saturated pores. Consistent with our previous work~\cite{Hamid2023,Hamid2020}, the flow is assumed steady, laminar, and incompressible.

Electromigration is neglected in the present formulation. This approximation applies under any of the following conditions:
\begin{itemize}[leftmargin=*]
    \item The fluid is a binary electroneutral electrolyte for which an apparent salt conservation equation can be derived from the Nernst--Planck equations \cite{Lai2011,Newman2004}.
    \item The solute is dilute and a sufficiently concentrated supporting electrolyte is present, such that the electric field $\mathbf{E}=-\nabla\phi$ is negligible and the solute transference number $t_i\equiv (z_i^2 m_i c_i)/(\sum_{j=1}^N z_j^2 m_j c_j)$ satisfies $t_i\ll1$. Here, $z_i$ and $m_i$ denote the charge number and mobility of species $i$, respectively.
    \item The solute is uncharged.
\end{itemize}
In a general multicomponent system, the diffusive flux of one species may depend on the concentration gradients of several species. Here, we adopt a pseudo-binary approximation \cite{Newman2009}, under which the diffusive flux of species $i$ follows Fick's law, $\jj_i^d=-D_i\nabla c_i$, with constant diffusion coefficient $D_i$. Under these assumptions, Eq.~\eqref{eq:1} reduces to the transient advection-diffusion equation,
\begin{equation}
    \frac{\partial c_i}{\partial t} + \nabla\cdot\left(\mathbf{u}c_i - D_i\nabla c_i\right) = 0. 
    \label{eq:2}
\end{equation}

Heterogeneous reaction at the solid/liquid interface, denoted by the subscript $s$, is described by a Robin-type boundary condition with a time-dependent equilibrium concentration,
\begin{equation}
    \widehat{\mathbf{n}}\cdot D_i\left.\nabla c_i\right|_s = k_i\left[c_{i,s,eq}(t) - c_{i,s}(t)\right], 
    \label{eq:3}
\end{equation}
where $\nhat$ is the unit normal directed from the fluid toward the solid phase [Fig.~\ref{fig:1}(b)], $\ki$ is the interfacial reaction-rate constant, and $\cs$ and $\cseq$ denote the surface and equilibrium concentrations of species $i$, respectively. This boundary condition provides a unified description of interfacial reactivity, with the non-reactive and perfectly reactive cases recovered as limiting conditions.

For a system containing a single dissolved reactant, Eqs.~\eqref{eq:2}--\eqref{eq:3} are sufficient to describe species transport. In multicomponent systems, however, the concentration fields may be coupled through electroneutrality and interfacial reaction stoichiometry. As a representative example, consider a binary electroneutral electrolyte containing a dissolved redox couple, $O$ and $R$, undergoing the outer-sphere electron-transfer reaction $O+e^-\rightleftharpoons R$ at the solid/liquid interface. As shown in Sec.~A of the Supporting Information (SI), the governing equations and interfacial boundary conditions for $c_O$ and $c_R$ decouple when $D_O=D_R=D$ and $k_O=k_R=k$, conditions satisfied by many redox systems \cite{Dagar2024,Gong2016,Yamamura2005,Sum1985a,Sum1985b}. Under these conditions, both species can be represented by a single concentration field. Accordingly, the species subscript $i$ is omitted hereafter, and the subsequent analysis is formulated in terms of a representative species.

    \section{\label{sec:Sec_03}Frequency-Domain Upscaling of Transient Transport}

To describe transient concentration evolution under steady flow, Eqs.~\eqref{eq:2}--\eqref{eq:3} are transformed into the frequency domain using the continuous-time Fourier transform. The resulting boundary-value problem for the Fourier-transformed concentration field $\tc(\xl,\omega)$ is
\begin{equation}
    \iu\omega\tc + \nabla\cdot\left(\uu\tc\right) = \nabla\cdot\left(D\nabla\tc\right), 
    \label{eq:4}
\end{equation}
\begin{equation}
    \nhat\cdot D\left.\nabla\tc\right|_s = k\left[\tc_{s,eq}(\omega) - \tc_s(\omega)\right]. 
    \label{eq:5}
\end{equation}
Here, the Fourier transform of a time-dependent quantity $Q(\xl,t)$ is defined as
\begin{equation}
    \mathcal{F}\left\{Q(\xl,t)\right\} = \overline{Q}(\xl,\omega) = \int_{-\infty}^{+\infty} Q(\xl,t)\,e^{-\iu\omega t}\,dt, 
    \label{eq:6}
\end{equation}
where $\iu=\sqrt{-1}$ is the imaginary unit and the angular frequency is related to the frequency $f$ through $\omega=2\pi f$ \cite{Hamid2023}.

We upscale the frequency-domain governing equation (GE) in Eq.~\eqref{eq:4} by volume averaging over the RVE. This requires separation among the pore, RVE, and system length scales, such that $\lP\ll\lRVE\ll\Lsys$ (see Fig.~\ref{fig:1}). Under this condition, the local volume average of a quantity $Q(\xl,\omega)$ over the fluid phase is defined as
\begin{equation}
    \avg{Q} = \frac{1}{\Vf}\int_{\Vf} Q\,dV. 
    \label{eq:7}
\end{equation}
Applying this averaging operator to Eq.~\eqref{eq:4} yields the governing equation for the local volume-averaged concentration field $\avg{\tc}(\Xg,\omega)$,
\begin{equation}
    \iu\omega\avg{\tc} + \avg{\nabla\cdot(\uu\tc)} = D\avg{\nabla\cdot(\nabla\tc)}. 
    \label{eq:8}
\end{equation}
The averaged advection and diffusion terms are then expressed in terms of volume-averaged quantities using the spatial averaging theorems \cite{Gray1993,GrayLee1977,Whitaker1969,Drew1971}:
\begin{equation}
    \avg{\nabla\xi} = \nabla\avg{\xi} + \frac{1}{\Vf}\int_s \nhat\,\xi\,dA, \label{eq:9}
\end{equation}
\begin{equation}
    \avg{\nabla\cdot\boldsymbol{\zeta}} = \nabla\cdot\avg{\boldsymbol{\zeta}} + \frac{1}{\Vf}\int_s \nhat\cdot\boldsymbol{\zeta}\,dA. 
    \label{eq:10}
\end{equation}
Applying Eqs.~\eqref{eq:9} and \eqref{eq:10} to Eq.~\eqref{eq:8} introduces interfacial surface integrals and gives
\begin{align}
    \iu\omega\avg{\tc} & + \nabla\cdot\avg{\uu\tc} + \frac{1}{\Vf}\int_s \nhat\cdot\uu\tc\,dA = D\nabla\cdot\nabla\avg{\tc} \notag 
    \\
    & + \frac{D}{\Vf}\nabla\cdot\int_s \nhat\,\tc\,dA + \frac{D}{\Vf}\int_s \nhat\cdot\nabla\tc\,dA. 
    \label{eq:11}
\end{align}
To separate macro-scale and pore-scale transport, the velocity and concentration fields are decomposed into average and deviation components,
\begin{equation}
    \uu(\xl) \equiv \avg{\uu}(\Xg) + \up(\xl), 
    \label{eq:12}
\end{equation}
\begin{equation}
    \tc(\xl,\omega) \equiv \avg{\tc}(\Xg,\omega) + \tc'(\xl,\omega). \label{eq:13}
\end{equation}
Here, $\avg{\tc}(\Xg,\omega)$ and $\avg{\uu}(\Xg)$ describe the macroscale fields, while $\tc'(\xl,\omega)$ and $\up(\xl)$ represent their pore-scale deviations. By definition,
\begin{equation}
    \avg{\up} = \avg{\tc'} = 0.
    \label{eq:14}
\end{equation}
Substituting Eqs.~\eqref{eq:12} and \eqref{eq:13} into Eq.~\eqref{eq:11} separates the macroscale contributions from the pore-scale deviations and yields
\begin{align}
    \iu\omega\avg{\tc} & + \nabla\cdot\avg{\uu}\avg{\tc} + \nabla\cdot\avg{\up\,\tc'} + \left\{\frac{1}{\Vf}\int_s \nhat\cdot\uu\avg{\tc}\,dA\right\} \notag 
    \\
    & + \frac{1}{\Vf}\int_s \nhat\cdot\uu\tc'\,dA = D\nabla\cdot\nabla\avg{\tc} \notag 
    \\
    & + \left\{\frac{D}{\Vf}\nabla\cdot\int_s \nhat\,\avg{\tc}\,dA\right\}  + \frac{D}{\Vf}\nabla\cdot\int_s \nhat\,\tc'\,dA \notag
    \\
    & + \left\{\frac{D}{\Vf}\int_s \nhat\cdot\nabla\avg{\tc}\,dA\right\} + \frac{D}{\Vf}\int_s \nhat\cdot\nabla\tc'\,dA. 
    \label{eq:15}
\end{align}
The terms enclosed in curly braces in Eq.~\eqref{eq:15} represent nonlocal contributions associated with the finite RVE size. Taylor-series expansion and order-of-magnitude analysis show that these terms vanish when the pore, RVE, and system scales are sufficiently separated \cite{Whitaker1999},
\begin{equation}
    \lRVE \gg \lP, \label{eq:16}
    \end{equation}
    \begin{equation}
    \left(\lRVE\right)^2 \ll \left(\Lsys\right)^2, 
    \label{eq:17}
\end{equation}
where $\Lsys$ is the smallest characteristic length scale of the porous system. The derivation of these inequalities is provided in Sec.~B of the SI and follows the corresponding analysis of Whitaker and co-workers \cite{Whitaker1967,Quintard1993,Quintard1994b,Quintard1994c}. Equation~\eqref{eq:15} therefore reduces to
\begin{align}
    & \iu\omega\avg{\tc} + \nabla\cdot\avg{\uu}\avg{\tc} + \nabla\cdot\avg{\up\,\tc'} + \frac{1}{\Vf}\int_s \nhat\cdot\uu\tc'\,dA \notag 
    \\
    & = D\nabla\cdot\nabla\avg{\tc} + \frac{D}{\Vf}\nabla\cdot\int_s \nhat\,\tc'\,dA \quad + \frac{D}{\Vf}\int_s \nhat\cdot\nabla\tc'\,dA. 
    \label{eq:18}
\end{align}
To obtain an equation for the deviation concentration field, Eq.~\eqref{eq:13} is substituted into Eq.~\eqref{eq:4}, yielding
\begin{equation}
    \iu\omega\avg{\tc} + \iu\omega\tc' + \nabla\cdot\uu\avg{\tc} + \nabla\cdot\uu\tc' = D\nabla\cdot\nabla\avg{\tc} + D\nabla\cdot\nabla\tc'. 
    \label{eq:19}
\end{equation}
Subtracting Eq.~\eqref{eq:18} from Eq.~\eqref{eq:19} and invoking incompressibility, $\nabla\cdot\uu=0$, gives the governing equation for $\tc'$,
\begin{align}
    \iu\omega\tc' & + \up\cdot\nabla\avg{\tc} + \uu\cdot\nabla\tc' - \frac{1}{\Vf}\nabla\cdot\int_{\Vf} \up\,\tc'\,dV \notag 
    \\
    & - \left\{\frac{1}{\Vf}\int_s \nhat\cdot\uu\tc'\,dA\right\} = D\nabla\cdot\nabla\tc' \notag 
    \\
    & - \frac{D}{\Vf}\nabla\cdot\int_s \nhat\,\tc'\,dA - \left\{\frac{D}{\Vf}\int_s \nhat\cdot\nabla\tc'\,dA\right\}. 
    \label{eq:20}
\end{align}
Equation~\eqref{eq:20} is obtained without specifying the velocity boundary condition at the solid/liquid interface and therefore applies to both slip and no-slip flows. The terms enclosed in curly braces represent corrections associated with finite RVE size. An ansatz for the deviation concentration field $\tc'(\xl,\omega)$ is introduced next to close the governing equation for $\avg{\tc}(\Xg,\omega)$.

\subsection{Closure Problems for a Periodic Unit Cell}

\begin{table*}[t!]
    \caption{Summary of the closure problems for $a(\xl,\omega)$ and $\bcv(\xl,\omega)$, including their governing equations, interfacial boundary conditions, periodicity conditions, and null-mean constraints. The spatially invariant quantities $\alpha$ and $\boldsymbol\beta$ are determined by enforcing the corresponding null-mean constraints. Here, $\mathbf{q}$ denotes any integer linear combination of the lattice vectors.}
    \label{tab:1}
    \squeezetable
    \begin{ruledtabular}
        \begin{tabular}{lllc} \\[-8pt]
            \textbf{Variable} & \textbf{Governing Equation} & \textbf{Boundary Condition} & \textbf{Constraint} \\[4pt]
            \hline\\[-4pt]
            $\vcenter{\hbox{$a(\xl,\omega)$}}$ &
            $\vcenter{\hbox{$\iu\omega a+\nabla\cdot\left(\uu a-D\nabla a\right)=\alpha$}}$ &
            $\vcenter{\hbox{\shortstack[l]{$\nhat\cdot D\left.\nabla a\right|_s+k\left.a\right|_s=k$\\[2pt]$a(\xl+\mathbf q,\omega)=a(\xl,\omega)$}}}$ &
            $\vcenter{\hbox{$\avg{a}=0$}}$ \\[20pt]
            $\vcenter{\hbox{$\bcv(\xl,\omega)$}}$ &
            $\vcenter{\hbox{$\iu\omega\bcv+\nabla\cdot\left(\uu\bcv-D\nabla\bcv\right)=-\up+\boldsymbol\beta$}}$ &
            $\vcenter{\hbox{\shortstack[l]{$\nhat\cdot D\left.\nabla\bcv\right|_s+k\left.\bcv\right|_s=-D\nhat$\\[2pt]$\bcv(\xl+\mathbf q,\omega)=\bcv(\xl,\omega)$}}}$ &
            $\vcenter{\hbox{$\avg{\bcv}=0$}}$ \\ [10pt]
        \end{tabular}
    \end{ruledtabular}
\end{table*}

To close the upscaled equation for $\avg{\tc}$, we seek a representation of $\tc'(\xl,\omega)$ in terms of the pore- and macroscale driving forces that generate concentration inhomogeneity within the RVE. The inhomogeneous terms in the boundary-value problem for $\tc'$ in Eq.~\eqref{eq:20}, particularly the forcing term $\up\cdot\nabla\avg{\tc}$, motivate the linear decomposition introduced below. The interfacial boundary condition in Eq.~\eqref{eq:5} is first rewritten using the decomposition in Eq.~\eqref{eq:13}:
\begin{equation}
    \nhat\cdot D\left.\nabla\left(\tc' + \avg{\tc}\right)\right|_s = k\left(\tc_{s,eq} - \avg{\tc}_s - \left.\tc'\right|_s\right). \label{eq:21}
\end{equation}
Rearranging Eq.~\eqref{eq:21} gives
\begin{equation}
    \nhat\cdot D\left.\nabla\tc'\right|_s + k\left.\tc'\right|_s = k\left(\tc_{s,eq} - \avg{\tc}_s\right) - \nhat\cdot D\left.\nabla\avg{\tc}\right|_s. 
    \label{eq:22}
\end{equation}
Following the approach of Parada and coworkers for systems without interfacial reaction \cite{ValdesParada2011,DavitQuintard2012,ValdesParada2012}, the corresponding Green's-function representation for the present reactive interface can be written as
\begin{align}
    \tc' & = \int_{\Vf} -\up\cdot\nabla\avg{\tc}\,G\,dV_\delta \notag 
    \\
    & + \int_S G\biggl[k\left(\tc_{s,eq}-\avg{\tc}\right) - \nhat\cdot D\nabla\avg{\tc}\biggr]dA_\delta, 
    \label{eq:23}
\end{align}
where the integrations are performed over the source location $\xl_\delta$ associated with the Green's function $G(\xl,\omega|\xl_\delta)$. Under the assumed separation of macro- and micro-scale length scales, $\nabla\avg{\tc}$ and $\tc_{s,eq}-\avg{\tc}$ may be treated as constants over the RVE, giving
\begin{align}
    \tc' & = \left(\tc_{s,eq}-\avg{\tc}\right)\int_S kG\,dV_\delta \notag 
    \\
    & + \nabla\avg{\tc}\cdot\left[\int_{\Vf} -\up\,G\,dV_\delta \int_S -\nhat\,D\,G\,dA_\delta\right]. 
    \label{eq:24}
\end{align}
Motivated by Eq.~\eqref{eq:24}, the deviation concentration field is represented by the ansatz
\begin{equation}
    \tc'(\xl,\omega) = a(\xl,\omega)\left(\tc_{s,eq}-\avg{\tc}\right) + \bcv(\xl,\omega)\cdot\nabla\avg{\tc}(\Xg,\omega). 
    \label{eq:25}
\end{equation}
Here, $a(\xl,\omega)$ is dimensionless and $\bcv(\xl,\omega)$ has dimensions of length. The first term represents concentration inhomogeneity driven by the interfacial reaction, while the second captures the response to the macroscale concentration gradient. Setting $\nabla\avg{\tc}=0$ in Eq.~\eqref{eq:25} gives $\tc'=a(\xl,\omega)(\tc_{s,eq}-\avg{\tc})$, showing that $a(\xl,\omega)$ isolates the reaction-driven contribution to the pore-scale concentration field. In contrast, $\bcv(\xl,\omega)$ describes the contribution associated with the macroscale concentration gradient. Because the closure variables depend only on the local pore-scale coordinate $\xl$, and because their governing equations and boundary conditions are spatially periodic, $a(\xl,\omega)$ and $\bcv(\xl,\omega)$ inherit the periodicity of the microstructure:
\begin{equation}
    a(\xl+\mathbf{q},\omega)=a(\xl,\omega),
    \label{eq:26}
\end{equation}
\begin{equation}
    \bcv(\xl+\mathbf{q},\omega)=\bcv(\xl,\omega). 
    \label{eq:27}
\end{equation}
Here $\mathbf{q}$ is any integer combination of the lattice vectors.

Because $\tc'$ is constructed from reaction-driven and gradient-driven contributions, the closure variables satisfy the null-mean conditions
\begin{equation}
    \avg{a} = \avg{\bcv} = 0 
    \label{eq:28}
\end{equation}
so that $\avg{\tc'}=0$. Substituting Eq.~\eqref{eq:25} into Eq.~\eqref{eq:20} yields
\begin{align}
    &\Biggl[\iu\omega a + \nabla\cdot\bigl(\uu a - D\nabla a\bigr) - \frac{1}{\Vf}\int_s \nhat\cdot\Bigl(\uu a \notag 
    \\
    & - D\nabla a\Bigr)dA\Biggr]\bigl(\tc_{s,eq}-\avg{\tc}\bigr) + \Biggl[\iu\omega\bcv + \nabla\cdot\bigl(\uu\bcv - D\nabla\bcv\bigr) \notag
    \\
    & + \up - \frac{1}{\Vf}\int_s \nhat\cdot\bigl(\uu\bcv - D\nabla\bcv\bigr)dA\Biggr]\cdot\nabla\avg{\tc} = 0,
    \label{eq:29}
\end{align}
where the scale-separation conditions in Eqs.~\eqref{eq:16} and \eqref{eq:17} are used to eliminate the divergence of certain integrals, as detailed in Sec.~C of the SI. The closure problem for $a(\xl,\omega)$ is obtained by considering the limit $\nabla\avg{\tc}=0$,
\begin{equation}
    \iu\omega a + \nabla\cdot\left(\uu a - D\nabla a\right) = \frac{1}{\Vf}\int_s \nhat\cdot\left(\uu a - D\nabla a\right)dA, 
    \label{eq:30}
\end{equation}
where incompressibility has additionally been invoked. Subtracting Eq.~\eqref{eq:30} from Eq.~\eqref{eq:29} then gives the governing equation for $\bcv(\xl,\omega)$,
\begin{equation}
    \iu\omega\bcv + \nabla\cdot\left(\uu\bcv - D\nabla\bcv\right) = -\up + \frac{1}{\Vf}\int_s \nhat\cdot\left(\uu\bcv - D\nabla\bcv\right)dA. \label{eq:31}
\end{equation}
The periodicity of $a(\xl,\omega)$ and $\bcv(\xl,\omega)$ makes the surface integrals on the right-hand sides of Eqs.~\eqref{eq:30} and \eqref{eq:31} spatially invariant. We therefore define $\alpha(\omega) \equiv (1/\Vf)\int_s \nhat\cdot\left(\uu a - D\nabla a\right)dA$ and $\boldsymbol\beta(\omega) \equiv (1/\Vf)\int_s \nhat\cdot\left(\uu\bcv - D\nabla\bcv\right)dA$, so that Eqs.~\eqref{eq:30} and \eqref{eq:31} reduce to
\begin{equation}
    \iu\omega a + \nabla\cdot\left(\uu a - D\nabla a\right) = \alpha, 
    \label{eq:32}
\end{equation}
\begin{equation}
    \iu\omega\bcv + \nabla\cdot\left(\uu\bcv - D\nabla\bcv\right) = -\up +\boldsymbol\beta.
    \label{eq:33}
\end{equation}
These equations remain nonlocal integro-differential equations because $\alpha$ and $\boldsymbol\beta$ depend on the corresponding solutions. Thus, for arbitrarily prescribed values of $\alpha$ and $\boldsymbol\beta$, Eqs.~\eqref{eq:32} and \eqref{eq:33} admit families of solutions. The null-mean conditions $\avg{a}=\avg{\bcv}=0$ select unique values of $\alpha$ and $\boldsymbol\beta$. Sec.~D of the SI establishes this uniqueness for impermeable systems, i.e., $\nhat\cdot\uu=0$ at the solid/liquid interface, by showing that the governing operator in Eqs.~\eqref{eq:32} and \eqref{eq:33} has a trivial null space. This ensures a unique solution for any prescribed $\alpha$ and $\boldsymbol\beta$. The unique values satisfying the null-mean constraints are then obtained by superposition, as detailed in Sec.~D of the SI. Accordingly, the remainder of this work is restricted to impermeable solid/liquid interfaces.

The boundary conditions for $a(\xl,\omega)$ and $\bcv(\xl,\omega)$ follow from Eq.~\eqref{eq:5}. For $a(\xl,\omega)$, substituting Eq.~\eqref{eq:13} into Eq.~\eqref{eq:5} and imposing the null-gradient condition $\nabla\avg{\tc}=0$ gives
\begin{equation}
    \nhat\cdot D\left.\nabla a\right|_s + k\left.a\right|_s = k. 
    \label{eq:34}
\end{equation}
Similarly, substituting Eq.~\eqref{eq:13} into Eq.~\eqref{eq:5} and using Eq.~\eqref{eq:34} gives
\begin{equation}
    \nhat\cdot D\left.\nabla\bcv\right|_s + k\left.\bcv\right|_s = -D\nhat. 
    \label{eq:35}
\end{equation}
The closure problems for $a(\xl,\omega)$ and $\bcv(\xl,\omega)$, together with their boundary conditions and constraints, are summarized in Table~\ref{tab:1}.

\subsection{Upscaled Transport Equations and Effective Transport Coefficients}

We now use the closure variables to derive the upscaled mass conservation equation and identify the corresponding effective transport coefficients. Substituting the ansatz of Eq.~\eqref{eq:25} into Eq.~\eqref{eq:18} gives
\begin{align}
    & \iu\omega\avg{\tc} + \nabla\cdot\Biggl[\avg{\uu}\avg{\tc} - \Bigl(\frac{D}{\Vf}\int_s \nhat\,a\,dA - \notag
    \\
    & \frac{1}{\Vf}\int_{\Vf} \up\,a\,dV\Bigr)\bigl(\tc_{s,eq}-\avg{\tc}\bigr)\Biggr] - \nabla\cdot D\Biggl[\bb{I} + \frac{1}{\Vf}\int_s \nhat\,\bcv\,dA \notag
    \\
    & - \frac{1}{D\Vf}\int_{\Vf} \up\,\bcv\,dV\Biggr]\cdot\nabla\avg{\tc} = -\frac{\tc_{s,eq}-\avg{\tc}}{\Vf}\int_s \nhat\cdot\bigl(\uu a \notag 
    \\
    & - D\nabla a\bigr)dA - \Biggl[\frac{1}{\Vf}\int_s \nhat\cdot\bigl(\uu\bcv - D\nabla\bcv\bigr)dA\Biggr]\cdot\nabla\avg{\tc}.
    \label{eq:36}
\end{align}
Here, $\bb{I}$ is the identity matrix. Because $\iu\omega\avg{\tc}$ represents the Fourier-transformed accumulation rate per unit fluid volume, Eq.~\eqref{eq:36} is an intrinsic-average transport equation. Multiplying by the porosity $\eps=\Vf/\VRVE$ and using $\nhat \cdot (\uu a)=\nhat \cdot (\uu \bcv)=0$ at impermeable interfaces gives the corresponding superficial-average form,
\begin{align}
    & \iu\omega\eps\avg{\tc} + \nabla\cdot\Biggl[\eps\avg{\uu}\avg{\tc} \notag \\
    & - \left(\frac{\eps D}{\Vf}\int_s \nhat\,a\,dA - \frac{\eps}{\Vf}\int_{\Vf} \up\,a\,dV\right)\bigl(\tc_{s,eq}-\avg{\tc}\bigr)\Biggr]  \notag 
    \\ &- \nabla\cdot D\Biggl[\eps\bb{I} + \frac{\eps}{\Vf}\int_s \nhat\,\bcv\,dA - \frac{\eps}{D\Vf}\int_{\Vf} \up\,\bcv\,dV\Biggr]\cdot\nabla\avg{\tc} \notag
    \\
    & = \frac{\eps D\bigl(\tc_{s,eq}-\avg{\tc}\bigr)}{\Vf}\int_s \nhat\cdot\bigl(\nabla a\bigr)dA  \notag
    \\
    & + \Biggl[\frac{\eps D}{\Vf}\int_s \nhat\cdot\bigl(\nabla\bcv\bigr)dA\Biggr]\cdot\nabla\avg{\tc}.
    \label{eq:37}
\end{align}
Inspection of the terms multiplying the macroscopic driving forces in Eq.~\eqref{eq:37} identifies four effective transport parameters:
\begin{align}
    \Deff = D\left[\eps\bb{I} + \frac{\eps}{\Vf}\int_s \nhat\,\bcv\,dA - \frac{\eps}{D\Vf}\int_{\Vf} \up\,\bcv\,dV\right], 
    \label{eq:38}
\end{align}
\begin{align}
    \Weff = \left(\frac{\eps D}{\Vf}\int_s \nhat\,a\,dA - \frac{\eps}{\Vf}\int_{\Vf} \up\,a\,dV\right)\left(\tc_{s,eq}-\avg{\tc}\right), \label{eq:39}
\end{align}
\begin{align}
    \heff = \frac{\eps D}{\sA\Vf}\int_s \nhat\cdot \nabla a\,dA, \label{eq:40}
\end{align}
\begin{align}
    \Reff = - \frac{\eps D}{\sA\Vf}\int_s \nhat\cdot \nabla\bcv \,dA. \label{eq:41}
\end{align}
Equation~\eqref{eq:37} can therefore be written as an advection-dispersion-reaction equation (ADRE) with two reaction-related terms on the right-hand side:
\begin{align}
    & \iu\omega\eps\avg{\tc} + \nabla\cdot\left[\eps\avg{\uu}\avg{\tc} - \Weff\right] - \nabla\cdot\left[\Deff\cdot\nabla\avg{\tc}\right] \notag 
    \\
    & = \sA\heff\left(\tc_{s,eq}-\avg{\tc}\right) - \sA\Reff\cdot\nabla\avg{\tc}. 
    \label{eq:42}
\end{align}
Here, $\sA$ is the surface area per unit volume of the porous medium, $\sA=A_s/\VRVE$. The effective dispersion tensor $\Deff$ is a rank-two tensor that describes the effective diffusive transport arising from molecular diffusion, pore geometry, and pore-scale velocity variations. The effective advection-suppression vector $\Weff$ describes the reduction in apparent advective transport caused by reaction-induced pore-scale concentration inhomogeneity, analogous to the frequency-domain parameter introduced in our previous work \cite{Hamid2023}. The effective mass-transfer coefficient $\heff$ characterizes the upscaled contribution of heterogeneous interfacial reaction. The reactivity-bias vector $\Reff$ describes the modification of the net reaction rate caused by macroscale concentration gradients.

The nondimensional forms of these effective parameters are
\begin{equation}
    \Dnd = \frac{\Deff}{D} = \eps\left[\bb{I} + \frac{1}{\Vf}\int_s \nhat\,\bcv\,dA - \frac{1}{D\Vf}\int_{\Vf} \up\,\bcv\,dV\right], 
    \label{eq:43}
\end{equation}
\begin{align}
    \Nnd & = \left(\frac{1}{\eps\avg{u}}\right)\frac{\Weff}{\left(\tc_{s,eq}-\avg{\tc}\right)} \notag 
    \\
    & = \frac{D}{\avg{u}\Vf}\int_s \nhat\,a\,dA - \frac{1}{\avg{u}\Vf}\int_{\Vf} \up\,a\,dV, 
    \label{eq:44}
\end{align}
\begin{equation}
    \Shnd = \left(\frac{\heff\,\lP}{D}\right) = \frac{\eps\lP}{\sA\Vf}\int_s \nhat\cdot \nabla a\,dA, 
    \label{eq:45}
\end{equation}
\begin{equation}
    \Rnd = \frac{\Reff}{D} = - \frac{\eps}{\sA\Vf}\int_s \nhat\cdot \nabla\bcv\,dA. 
    \label{eq:46}
\end{equation}
The dispersion tensor $\Dnd$ has received the greatest attention among these four coefficients in the existing literature. While Parada and Ramirez~\cite{ValdesParada2011} and Moyne~\cite{Moyne1997} introduced dynamic dispersion tensors for non-reactive systems, the present work derives a frequency-dependent dispersion tensor that simultaneously accounts for advection, diffusion, and heterogeneous interfacial reaction. Following the interpretation of Whitaker and co-workers \cite{Whitaker1967,Quintard2006,Ryan1980}, the surface and volume integrals in Eq.~\eqref{eq:43} are identified with the tortuosity tensor $\tortu$ and hydrodynamic dispersion tensor $\hdisp$, respectively. Extending these quantities to the frequency domain gives
\begin{equation}
    \tortu = \frac{1}{\Vf}\int_s \nhat\,\bcv\,dA, \label{eq:47}
\end{equation}
\begin{equation}
    \hdisp = -\frac{1}{D\Vf}\int_{\Vf} \up\,\bcv\,dV. \label{eq:48}
\end{equation}
The definition of $\tortu$ in Eq.~\eqref{eq:47} provides insight into the longstanding debate over whether tortuosity is an intrinsic geometric property of porous media or a process-dependent transport characteristic \cite{Ghanbarian2013,Holzer2023}. Equation~\eqref{eq:47} shows that tortuosity depends on interfacial reactivity because the closure variable $\bcv(\xl,\omega)$ is governed by the interface-specific boundary condition in Eq.~\eqref{eq:35}. In the limit $\Da\to\infty$, corresponding to perfectly reactive interfaces, the tortuosity tensor becomes exactly zero. Physically, tortuosity reflects the effect of solid boundaries on diffusive transport through the pore space. As the interface becomes highly reactive, however, it acts as a solute source or sink rather than solely as a barrier to diffusion. The tortuosity contribution therefore vanishes in the perfectly reactive limit. Equation~\eqref{eq:48} further shows that the hydrodynamic dispersion tensor $\hdisp$ depends on both the pore geometry and the pore-scale concentration and velocity fields represented by $\bcv$. Its dependence on $D$ reflects the role of transverse molecular diffusion in relaxing the concentration gradients generated by velocity variations.

The second effective transport parameter, $\Nnd$ in Eq.~\eqref{eq:44}, quantifies the deviation of the net advective transport from the product of the mean velocity and mean concentration. We refer to this effect as advection suppression because reaction-induced concentration inhomogeneity causes the apparent advective transport of reacted solute to differ from that predicted using the bulk mean concentration alone. Accordingly, $\Nnd$ vanishes in the non-reactive limit. The surface and volume integrals in Eq.~\eqref{eq:44} are analogous to those defining the tortuosity and hydrodynamic dispersion tensors in Eqs.~\eqref{eq:47} and \eqref{eq:48}, but involve $a(\xl,\omega)$ rather than $\bcv(\xl,\omega)$. In the limit $\Da\to\infty$, the boundary condition in Eq.~\eqref{eq:34} causes the surface contribution to vanish. Under the null-gradient condition $\nabla\avg{\tc}=0$, the remaining volume integral captures the covariance between the pore-scale velocity and concentration fields, $\mathrm{covar}(\uu,\tc)=\avg{\uu\tc}-\avg{\uu}\avg{\tc}$. A specific form of $\Nnd$ was previously introduced in our earlier work \cite{Hamid2023} without the volume-averaging framework developed here. Section~E of the SI shows that the present formulation reduces to that earlier definition in the limit $\Da\to\infty$ with $\nabla\avg{\tc}=0$.

The third coefficient, the spectral Sherwood number $\Shnd$ defined by Eq.~\eqref{eq:45}, characterizes the transient interfacial mass-transfer response of the porous medium \cite{Hamid2023}. Unlike the classical Sherwood number, which describes steady-state transport, $\Shnd$ is frequency-dependent and captures the response of interfacial mass transfer to time-dependent forcing. Because it depends on $a(\xl,\omega)$ rather than $\bcv(\xl,\omega)$, it is determined by pore-scale concentration gradients associated with the reactive interface and is independent of the macroscale concentration gradient. Physically, $\Shnd$ measures the efficiency of diffusive transport to and from reactive surfaces under transient forcing. At short time scales, the diffusion boundary layer evolves dynamically rather than remaining fixed as in classical film theory. Consequently, $\Shnd$ increases at high frequencies, extending the film-based description of mass transfer to transient conditions. Although introduced previously in Ref.~\cite{Hamid2023}, the present derivation extends its use to systems with nonzero $\nabla\avg{\tc}$. Section~F of the SI further shows that the present expression reduces to the earlier formulation in the appropriate limit.

The fourth coefficient, the reactivity-bias vector $\Rnd$ defined by Eq.~\eqref{eq:46}, quantifies how a macroscale concentration gradient modifies interfacial reaction through flow-induced redistribution of solute. To the best of our knowledge, this is the first transport parameter of this type derived through a volume-averaging framework. Unlike dispersion, which characterizes solute spreading, $\Rnd$ quantifies the effect of macroscale concentration gradients on the net interfacial reaction rate in the presence of flow. Diffusion tends to homogenize the concentration field across streamlines, whereas advection transports solute preferentially along regions of different velocity. As a result, reactive interfaces experience spatially nonuniform transport, with some regions receiving enhanced solute flux and others receiving reduced flux. The reactivity-bias vector provides a macroscopic measure of this imbalance. It vanishes in the absence of flow, as shown later.

The four nondimensional transport coefficients defined by Eqs.~\eqref{eq:43}--\eqref{eq:46} can also be interpreted as transfer functions that relate the relevant driving forces to their corresponding effective fluxes, as illustrated in Fig.~\ref{fig:2}(b). In systems theory, a transfer function describes the frequency-dependent relation between an input and an output. Here, the relevant driving forces are the macroscale concentration gradient $\nabla\avg{\tc}$ and the reactive driving force $\tc_{s,eq}-\avg{\tc}$, while the corresponding outputs are the effective diffusive, advective, and reactive fluxes appearing in Eq.~\eqref{eq:42}. For example, the dispersion tensor $\Dnd$ relates $\nabla\avg{\tc}$ to the effective diffusive flux through $\avg{\jj^d}_{eff}=D\,\Dnd\cdot\nabla\avg{\tc}$. Because $\Dnd$ depends on $\omega$, the magnitude and phase of the flux response depend on the frequency of the driving force. Similarly, $\Nnd$ describes the reaction-induced modification of effective advection, $\Shnd$ characterizes interfacial mass transfer driven by the reactive concentration difference, and $\Rnd$ describes the contribution of the macroscale concentration gradient to the effective reactive flux. Thus, the four transfer functions quantify how pore-scale transport and interfacial reactivity modify the corresponding driving forces before they appear as macroscale fluxes.

Using these transfer functions, the upscaled mass-conservation equation in the frequency domain becomes
\begin{align}
    & \eps\iu\omega\avg{\tc} + \nabla\cdot\Bigl(\eps\avg{\uu}\avg{\tc} - \eps\avg{u}\Nnd\bigl(\tc_{s,eq}-\avg{\tc}\bigr)\Bigr) \notag 
    \\
    & - D\nabla\cdot\bigl(\Dnd\cdot\nabla\avg{\tc}\bigr) = \frac{\sA D}{\lP}\Shnd\bigl(\tc_{s,eq}-\avg{\tc}\bigr) \notag 
    \\
    & - \sA D\Rnd\cdot\nabla\avg{\tc}.
    \label{eq:49}
\end{align}
Applying the inverse Fourier transform to Eq.~\eqref{eq:49} gives the corresponding time-domain upscaled mass-conservation equation,
\begin{align}
    \eps\frac{\partial\avg{c}}{\partial t} & + \nabla\cdot\Bigl(\eps\avg{\uu}\avg{c} - \eps\avg{u}\,\mathcal{F}^{-1}\bigl\{\Nnd\bigl(\tc_{s,eq}-\avg{\tc}\bigr)\bigr\}\Bigr) \notag 
    \\
    & - D\nabla\cdot\mathcal{F}^{-1}\bigl\{\Dnd\cdot\nabla\avg{\tc}\bigr\} = \frac{\sA D}{\lP}\mathcal{F}^{-1}\bigl\{\Shnd \notag
    \\
    & \quad \bigl(\tc_{s,eq} -\avg{\tc}\bigr)\bigr\} - \sA D\,\mathcal{F}^{-1}\bigl\{\Rnd\cdot\nabla\avg{\tc}\bigr\}.
    \label{eq:50}
\end{align}

Equation~\eqref{eq:50} also shows how the frequency-dependent formulation captures the memory inherent in transient upscaled transport. In the time domain, each effective flux is obtained from the inverse Fourier transform of a product of a frequency-dependent transfer function and its corresponding driving force. For example, the effective diffusive flux depends on $\mathcal{F}^{-1}\{\Dnd\cdot\nabla\avg{\tc}\}$, which, by the convolution theorem, can be written as a temporal convolution between the memory kernel associated with $\Dnd$ and the history of the macroscale concentration gradient. Analogous convolution relations apply to $\Nnd$, $\Shnd$, and $\Rnd$. The resulting flux at time $t$ therefore depends not only on the instantaneous driving force but also on its history. Physically, this memory reflects the finite time required for the pore-scale concentration field to adjust to changes in the macroscale gradient or reactive driving force, with a characteristic time of order $\lP^2/D$. Because each transfer function in Eq.~\eqref{eq:50} is derived directly from the pore-scale closure problem, its corresponding memory kernel follows from inverse Fourier transformation rather than from an assumed functional form, in contrast to delay-diffusion or exponential-relaxation models~\cite{Smith1981,Smith1982}.

In the following sections, analytical expressions for these frequency-dependent transport coefficients are obtained by solving the closure problems summarized in Table~\ref{tab:1} for Poiseuille flow between parallel plates and through circular tubes across the full range of interfacial reactivity, from $\Da=0$ to $\Da\to\infty$. The resulting transfer functions are then used with the inverse fast Fourier transform (iFFT) to obtain the corresponding upscaled time-domain response governed by Eq.~\eqref{eq:50}. The formulation is also applied to a model porous medium consisting of a regular array of square rods under cross flow, evaluated in the non-reactive ($\Da=0$) and perfectly reactive ($\Da\to\infty$) limits.

    \section{\label{sec:Sec_04}Pore-Scale Frequency-Dependent Transport in Flow between Parallel Plates and through Circular Tubes}

\begin{figure}[t!]
    \centering
    \includegraphics[width=0.9\linewidth]{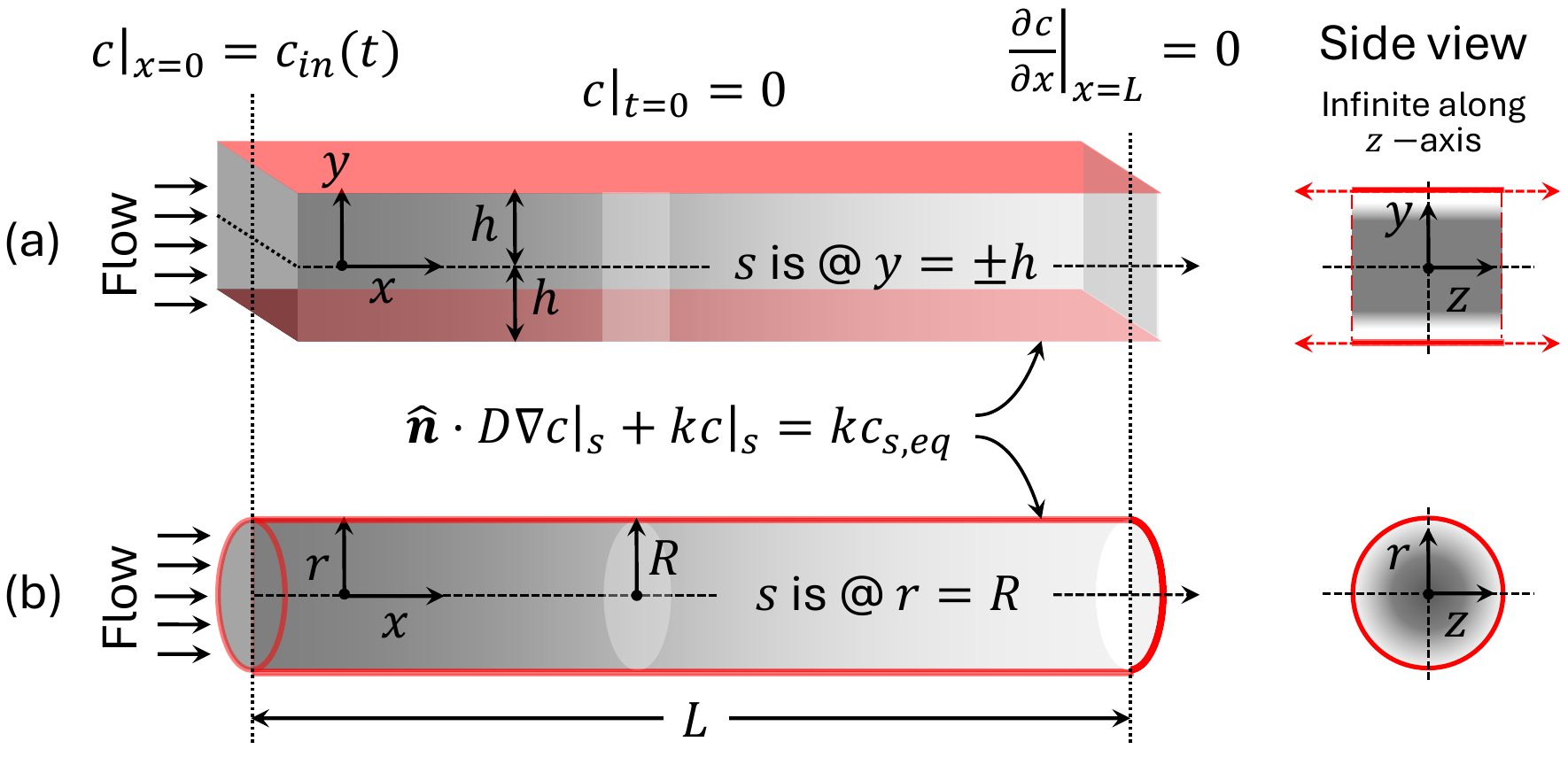}
    \caption{Laminar, fully developed Poiseuille flow (a) between parallel plates with half-gap $h$ and (b) through a circular tube with radius $R$. Both geometries have length $L$, with the solid/liquid interface (red) denoted by $s$. The fluid is initially solute-free, $c|_{t=0}=0$, and both systems are subject to the interfacial Robin boundary condition $\nhat\cdot D\nabla c|_s+kc|_s=kc_{s,eq}$, a time-varying inlet concentration $c|_{x=0}=c_{in}(t)$, and a zero-gradient outlet condition $\partial c/\partial x|_{x=L}=0$. The right panels show cross-sectional views along the flow ($x$) direction; the parallel-plate geometry is assumed infinite along the $z$-axis.}
    \label{fig:3}
\end{figure}

\begin{figure*}[t!]
    \centering
    \includegraphics[width=0.75\linewidth]{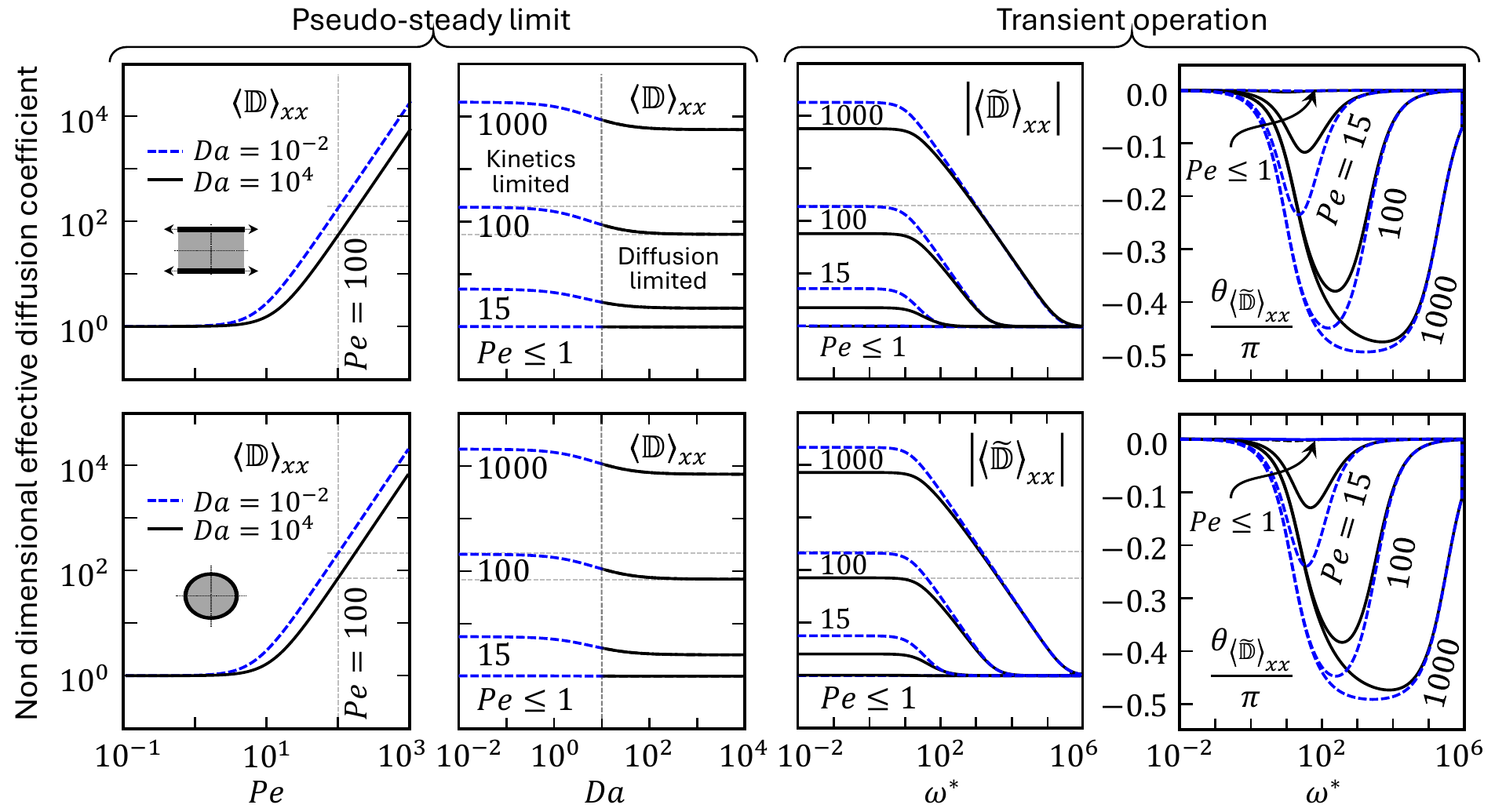}
    \caption{Longitudinal dispersion coefficient for flow between parallel plates (top row) and through a circular tube (bottom row) under pseudo-steady (first two columns) and transient (last two columns) conditions. The first column shows the $\Pe$ dependence of $\avg{\Dbb}_{xx}$ for $\Da=10^{-2}$ and $\Da=10^4$, while the second column shows its $\Da$ dependence for different $\Pe$. For $\Pe\le1$, the curves overlap. The third and fourth columns show the magnitude and phase, respectively, of the frequency-dependent dispersion coefficient $\Dxx$ as functions of $\wstar$ for different $\Pe$. Dashed and solid lines correspond to $\Da=10^{-2}$ and $\Da=10^4$, respectively. Light-dotted horizontal lines indicate the corresponding pseudo-steady values.}
    \label{fig:4}
\end{figure*}

\begin{figure}[t!]
    \centering
    \includegraphics[width=0.9\linewidth]{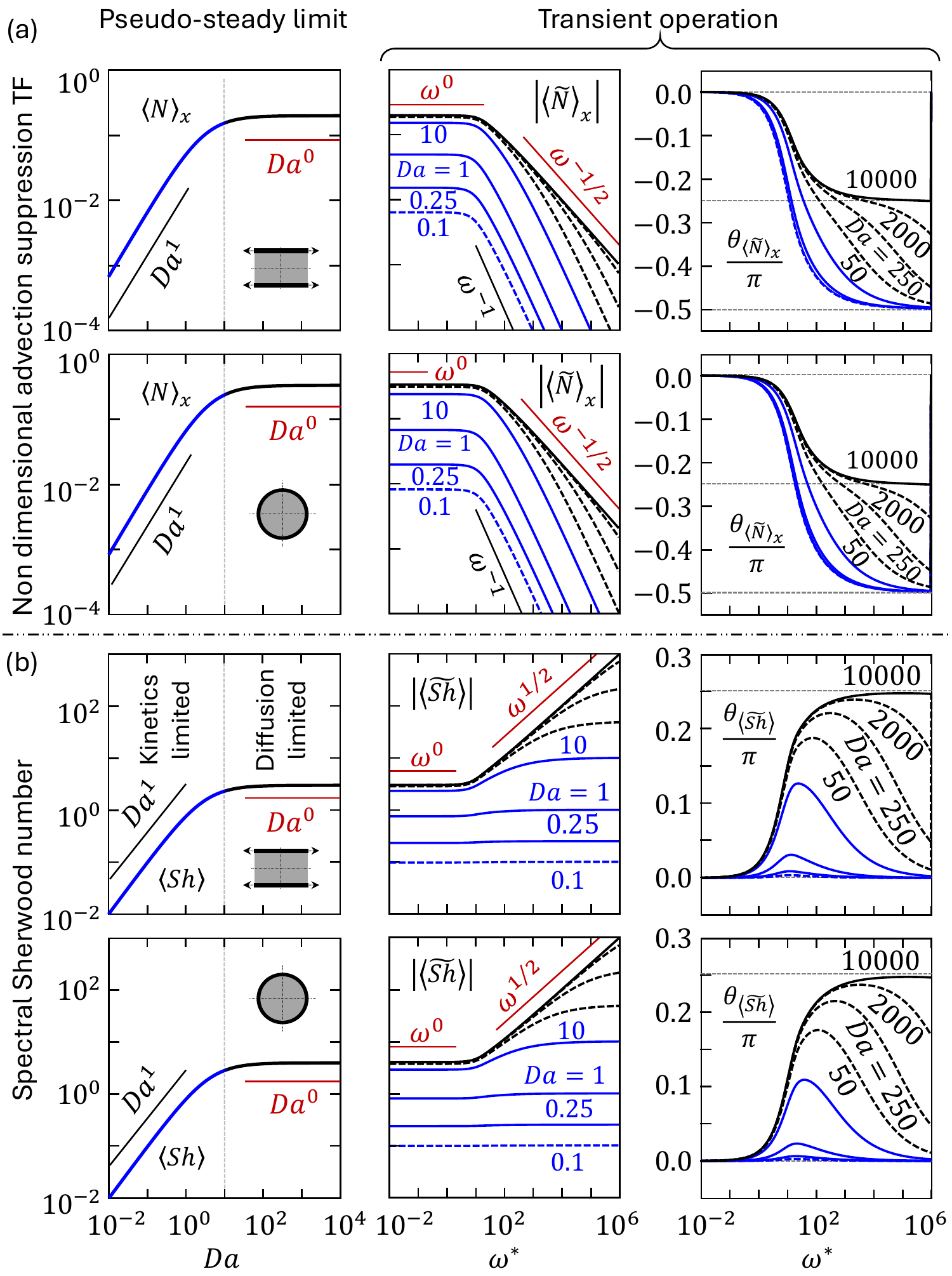}
    \caption{Variation of (a) the advection-suppression transfer function and (b) the spectral Sherwood number under pseudo-steady (first column) and transient (second and third columns) conditions. In both (a) and (b), the top and bottom rows correspond to parallel plates and a circular tube, respectively. The first column shows the transition between the kinetics-limited and diffusion-limited regimes as $\Da$ varies. The second and third columns show the magnitude and phase, respectively, of the frequency-dependent response for different $\Da$. Blue and black curves correspond to the kinetics-limited and diffusion-limited regimes, respectively. The $\Da$ values are distributed between the second- and third-column panels for clarity.}
    \label{fig:5}
\end{figure}

\begin{table*}[t!]
    \caption{Analytical expressions for the dispersion coefficient $\avgt{\Dbb}_{xx}$, advection-suppression coefficient $\avgt{N}_x$, spectral Sherwood number $\Shnd$, and reactivity-bias coefficient $\avgt{R}_x$ under pseudo-steady (PSL) and frequency-dependent (omni-temporal) conditions for Poiseuille flow between parallel plates and through a circular tube.}
    \label{tab:2}
    \squeezetable
    \renewcommand{\arraystretch}{1.4}
    \begin{ruledtabular}
        \begin{tabular}{l@{\hspace{5pt}}l@{\hspace{12pt}}l@{\hspace{12pt}}l}
            & \textbf{Operation} & \textbf{Parallel plates} & \textbf{Circular tube} \\ [4pt]
            \hline \\ [-10pt]
            \multirow{2}{*}[-3pt]{\shortstack[l]{Longitudinal\\dispersion\\coefficient $\avgt{\Dbb}_{xx}$}}
                & PSL & $\avg{\Dbb}_{xx} = 1 + \dfrac{1}{175}\left(\dfrac{\Da+10}{\Da+3}\right)\Pe^2$
                & $\avg{\Dbb}_{xx} = 1 + \dfrac{1}{144}\left(\dfrac{\Da+12}{\Da+4}\right)\Pe^2$ \\[10pt]
                & Omni-Temporal & $\avgt{\Dbb}_{xx} = 1 + \left(\dfrac{\Gamma_1}{5\xi^4\Gamma_2}\right)\Pe^2$
                & $\avgt{\Dbb}_{xx} = 1 + \left(\dfrac{\Gamma_4}{3\xi^4\Gamma_5}\right)\Pe^2$ \\ [10pt]
            \hline \\ [-10pt]
            \multirow{2}{*}[-4pt]{\shortstack[l]{Advection\\suppression\\TF $\avgt{N}_x$}}
                & PSL & $\avg{N}_x = \dfrac{\Da}{5(\Da+3)}$
                & $\avg{N}_x = \dfrac{\Da}{3(\Da+4)}$ \\[10pt]
                & Omni-Temporal & $\avgt{N}_x = \dfrac{\Da\bigl[(\xi^2+3)\sinh\xi-3\xi\cosh\xi\bigr]}{\xi^2\Gamma_2}$
                & $\avgt{N}_x = \dfrac{2\Da\bigl[(\xi^2+8)\,I_1(\xi)-4\xi\,I_0(\xi)\bigr]}{\xi^2\Gamma_5}$ \\ [10pt]
            \hline \\ [-10pt]
            \multirow{2}{*}[-5pt]{\shortstack[l]{Spectral\\Sherwood\\number $\Shnd$}}
                & PSL & $\avg{Sh} = \dfrac{3\Da}{3+\Da}$
                & $\avg{Sh} = \dfrac{4\Da}{4+\Da}$ \\[10pt]
                & Omni-Temporal & $\Shnd = \dfrac{\Da\,\xi^2\sinh\xi}{\Gamma_2}$
                & $\Shnd = \dfrac{\Da\,\xi^2\,I_1(\xi)}{\Gamma_5}$ \\ [10pt]
            \hline \\ [-10pt]
            \multirow{2}{*}[-3pt]{\shortstack[l]{Reactivity\\bias\\vector $\avgt{R}_x$}}
                & PSL & $\avg{R}_x = \Pe \left[\dfrac{\Da}{5(\Da+3)}\right]$
                & $\avg{R}_x = \Pe \left[\dfrac{\Da}{6(\Da+4)}\right]$ \\[10pt]
                & Omni-Temporal & $\avgt{R}_x = \Pe \left[\dfrac{\Gamma_2-3\Gamma_3}{2\Gamma_2} - \dfrac{3}{\xi^2}\right]$
                & $\avgt{R}_x = \Pe \left[\dfrac{(\Da+4)\,I_1(\xi)}{\Gamma_5} - \dfrac{4}{\xi^2}\right]$ \\ [10pt]
            \hline \\ [-10pt]
            \multirow{8}{*}[-10pt]{Here,}
                & \multicolumn{3}{l}{$\wstar = \dfrac{\omega(\lP)^2}{D}$; \quad $\xi=\sqrt{\iu\wstar}$; \quad $\Pe = \dfrac{\avg{u}\lP}{D}$; \quad $\Da = \dfrac{k\lP}{D}$.} \\[6pt]
                & \multicolumn{3}{l}{$\Gamma_1 = \bigl(\Da\xi^3+15\Da\xi+45\xi\bigr)\cosh\xi - \bigl(6\Da\xi^2+15\Da-\xi^4+15\xi^2+45\bigr)\sinh\xi$;} \\[4pt]
                & \multicolumn{3}{l}{$\Gamma_2 = \Da\,\xi\cosh\xi - \bigl(\Da-\xi^2\bigr)\sinh\xi$;} \\[4pt]
                & \multicolumn{3}{l}{$\Gamma_3 = \dfrac{\Da\,\xi}{3}\cosh\xi - \left(\Da+2-\dfrac{\xi^2}{3}\right)\sinh\xi$;} \\[4pt]
                & \multicolumn{3}{l}{$\Gamma_4 = \bigl(\Da\xi^3+24\Da\xi+96\xi\bigr)\,I_0(\xi) - \bigl(8\Da\xi^2+48\Da-\xi^4+24\xi^2+192\bigr)\,I_1(\xi)$;} \\[4pt]
                & \multicolumn{3}{l}{$\Gamma_5 = \Da\,\xi\,I_0(\xi) - \bigl(2\Da-\xi^2\bigr)\,I_1(\xi)$;} \\[6pt]
                & \multicolumn{3}{l}{$\lP$ represents $h$ for flow between parallel plates, and $R$ for flow through circular tube;} \\[4pt]
                & \multicolumn{3}{l}{$I_n(\xi)$ is the modified Bessel function of first kind; order $n$ and argument $\xi$.} \\[4pt]
        \end{tabular}
    \end{ruledtabular}
\end{table*}

\begin{figure}[t!]
    \centering
    \includegraphics[width=0.9\linewidth]{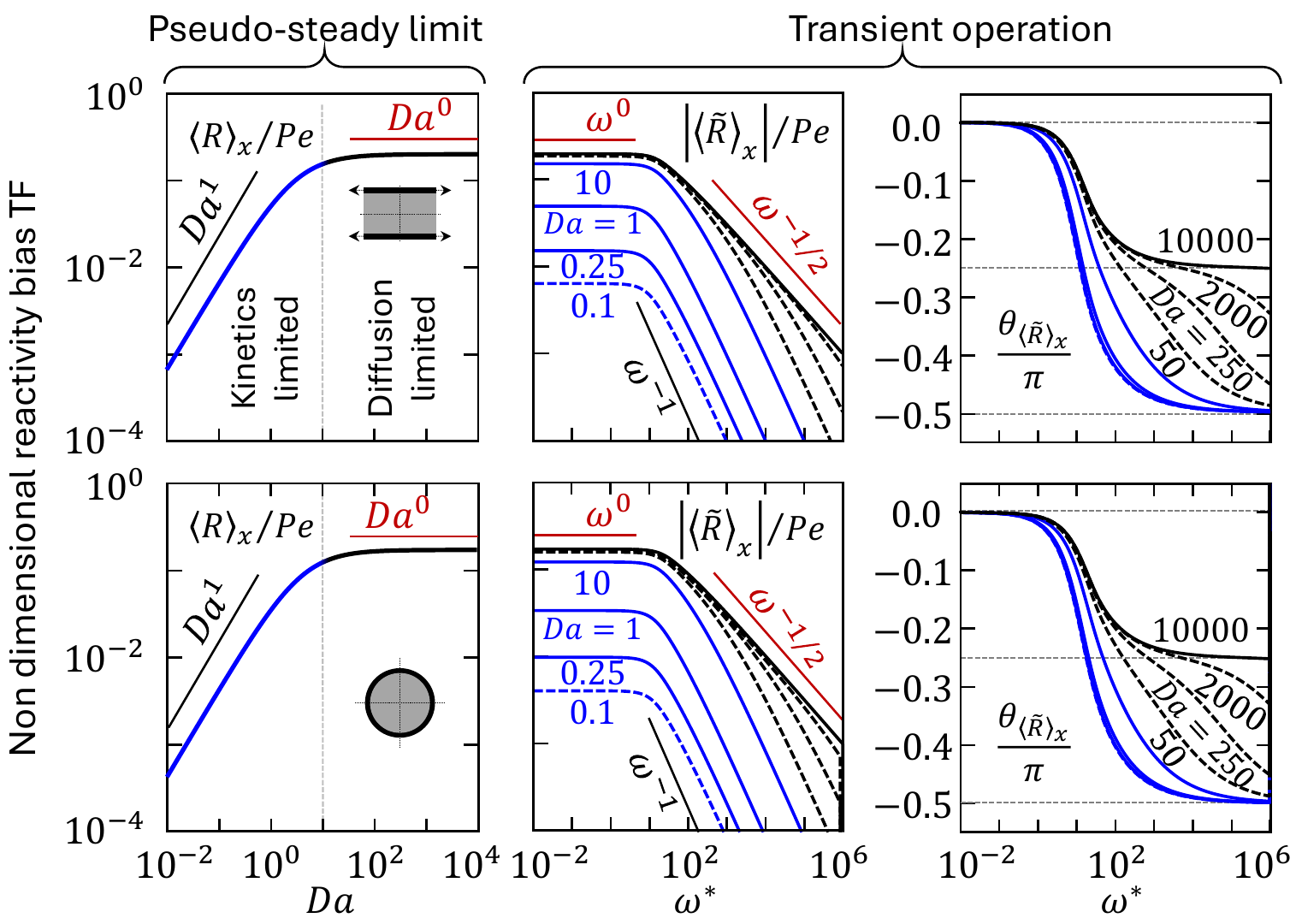}
    \caption{Variation of the reactivity-bias transfer function under pseudo-steady (first column) and transient (second and third columns) conditions. The top and bottom rows correspond to parallel plates and a circular tube, respectively. The first column shows the transition between the kinetics-limited and diffusion-limited regimes as $\Da$ varies. The second and third columns show the magnitude and phase, respectively, of the frequency-dependent response for different $\Da$. Blue and black curves correspond to the kinetics-limited and diffusion-limited regimes, respectively. The $\Da$ values are distributed between the second- and third-column panels for clarity.}
    \label{fig:6}
\end{figure}

Although the present framework applies to arbitrary periodic porous microstructures, analytical solutions are derived here for the canonical cases of fully developed flow between parallel plates and through circular tubes (see Fig.~\ref{fig:3}). The flow is assumed steady, laminar and unidirectional, with no-slip boundary conditions. The corresponding velocity profiles are $u_x=(3/2)\avg{u_x}[1-(y/h)^2]$ for parallel plates and $u_x=2\avg{u_x}[1-(r/R)^2]$ for circular tubes. The full range of interfacial reactivity, from $\Da=0$ to $\Da\to\infty$, is considered. For the axial transport problem considered here, the closure field is written as $\bcv(\xl,\omega)=b_x\widehat{\mathbf{i}}$, while $a$ and $b_x$ depend only on the transverse coordinate $y$ for parallel plates or the radial coordinate $r$ for circular tubes. The closure problems summarized in Table~\ref{tab:1} therefore reduce to ordinary differential equations (ODEs). The corresponding pseudo-steady equations are recovered by setting $\omega=0$. The reduced ODEs and boundary conditions for both pseudo-steady and frequency-dependent transport are summarized in Table~G.1, and their analytical solutions are given in Table~G.2 in Sec.~G of the SI.

The analytical solutions in Table~G.2 (Sec.~G of the SI) are substituted into Eqs.~(G.5)--(G.12), which are the reduced forms of Eqs.~\eqref{eq:43}--\eqref{eq:46}, to obtain the longitudinal dispersion coefficient $\Dxx$, advection-suppression coefficient $\Nx$, spectral Sherwood number $\Shnd$, and longitudinal reactivity-bias coefficient $\Rx$. Their frequency-independent forms in the pseudo-steady limit are denoted without the {\em ``tilde''}. The resulting analytical expressions are summarized in Tab.~\ref{tab:2}. Figures~\ref{fig:4}--\ref{fig:6} show their dependence on the P\'eclet number $\Pe=\avg{u}\lP/D$, Damk\"ohler number $\Da=k\lP/D$, and nondimensional frequency $\wstar=\omega(\lP)^2/D$, where $\lP=h$ for parallel plates and $\lP=R$ for a circular tube. The following subsections examine the behavior of each coefficient and its physical significance.

The transverse coordinate ($y$ or $r$) is resolved explicitly at the pore scale, while the upscaled transport problem is formulated only along the mean-flow direction. Thus, the intrinsic average $\avg{\tc}$ is taken to vary only along the axial coordinate $x$, and no independent transverse transport equation is considered in these one-dimensional single-pore reductions. Consequently, the quantities defined and reported here are the longitudinal dispersion coefficient $\Dxx$, the longitudinal component of the advection-suppression vector $\Nx$, the scalar spectral Sherwood number $\Shnd$, and the longitudinal component of the reactivity-bias vector $\Rx$.

\subsection{Frequency-Dependent Dispersion}

The first column of Fig.~\ref{fig:4} shows the variation of $\avg{\Dbb}_{xx}$ with $\Pe$ for two representative values of $\Da$: $\Da=10^{-2}$, corresponding to weak interfacial reaction, and $\Da=10^4$, corresponding to a nearly perfect reactive interface. In the limit of weak flow ($\Pe<1$), $\avg{\Dbb}_{xx}$ approaches the molecular-diffusion limit. As $\Pe$ increases, hydrodynamic dispersion becomes increasingly important, and for $\Pe\geq10$, $\avg{\Dbb}_{xx}$ follows the relation $\avg{\Dbb}_{xx}=1+\kappa\Pe^2$, where $\kappa$ depends on geometry and interfacial reactivity, as shown in Tab.~\ref{tab:2}. In the non-reactive limit ($\Da=0$), $\kappa$ is $2/105$ and $1/48$ for parallel plates and circular tubes, respectively, recovering the Taylor--Aris scaling. In the perfectly reactive limit ($\Da\to\infty$), these values decrease to $1/175$ and $1/144$, respectively, showing that interfacial reaction suppresses hydrodynamic dispersion.

The second column of Fig.~\ref{fig:4} shows the dependence of $\avg{\Dbb}_{xx}$ on $\Da$ for different $\Pe$. For each $\Pe$, the dispersion coefficient transitions between two limiting behaviors: a kinetics-limited regime at small $\Da$ and a diffusion-limited regime at large $\Da$. The dashed blue and solid black line segments identify these regimes for $\Da<10$ and $\Da>10$, respectively. The corresponding limiting values agree with those shown in the first column. This correspondence is highlighted for $\Pe=100$ using the light-dotted horizontal lines.

The last two columns show Bode plots of $\Dxx$, i.e., the frequency dependence of its magnitude and phase, for the two limiting $\Da$ cases at four values of $\Pe$. Throughout this study, the magnitude and phase of a complex-valued parameter $Q=Q_{real}+\iu Q_{imag}$ are calculated as $|Q|=\sqrt{Q_{real}^2+Q_{imag}^2}$ and $\theta_Q=\tan^{-1}(Q_{imag}/Q_{real})$, respectively. The light-dotted horizontal lines show the corresponding PSL values approached as $\wstar\to0$. As $\wstar$ increases, $|\Dxx|$ progressively approaches the purely diffusive limit, $|\Dxx|_{\omega\to\infty}\to1$. At sufficiently high frequency, the characteristic transverse diffusion length becomes small relative to the pore dimension, reducing the interaction between transverse diffusion and the nonuniform velocity field and thereby suppressing hydrodynamic dispersion. The corresponding phase varies non-monotonically with frequency: $\theta_{\Dxx}$ initially departs from zero, reaches a minimum at intermediate $\wstar$, and subsequently returns toward $\theta_{\Dxx}\to0$ as $\wstar\to\infty$. This phase response reflects the competition between advective distortion and transverse diffusive relaxation of the concentration field and provides a characteristic signature of pre-asymptotic hydrodynamic dispersion that is absent from pseudo-steady or asymptotic models.

\subsection{Frequency-Dependent Advection Suppression by Interfacial Reactions}

Because the interfaces in both geometries are aligned with the flow direction, $\avg{N}_x$ and $\Nx$ are independent of $\Pe$, as evident from the analytical expressions in Tab.~\ref{tab:2}. Both coefficients, however, depend on $\Da$. As shown in Fig.~\ref{fig:5}(a), two distinct regimes emerge: a kinetics-limited regime for $\Da<10$ and a diffusion-limited regime for $\Da>10$. At sufficiently small $\Da$, $\avg{N}_x$ increases linearly with $\Da$, reflecting its direct dependence on the interfacial reaction-rate constant when reactions are rate-limiting. The corresponding asymptotic slopes are $1/15$ and $1/12$ for parallel plates and circular tubes, respectively. As $\Da$ increases, the response becomes progressively less sensitive to $\Da$ as interfacial transport becomes diffusion-limited. Consequently, $\avg{N}_x$ approaches limiting values of $1/5$ and $1/3$ for parallel plates and circular tubes, respectively. These values agree exactly with our previously reported PSL results for the perfectly reactive limit \cite{Hamid2023}.

The Bode plots in the right two columns of Fig.~\ref{fig:5}(a) reveal corresponding differences in the high-frequency response. For $\wstar>10$, the kinetics-limited response follows $|\Nx|\sim\omega^{-1}$ with $\theta_{\Nx}\to\pi/2$, characteristic of a capacitive-type response. In the diffusion-limited regime, $|\Nx|\sim\omega^{-1/2}$ with $\theta_{\Nx}\to\pi/4$, corresponding to the classical semi--infinite Warburg response~\cite{Huang2018}. Thus, increasing $\Da$ changes the high-frequency response from kinetics-controlled to diffusion-controlled behavior.

\subsection{Frequency-Dependent Interfacial Mass Transfer: The Spectral Sherwood Number}

Both $\avg{Sh}$ and $\Shnd$ are independent of $\Pe$, as evidenced from the analytical expressions in Table~\ref{tab:2}. Their dependence on $\Da$ is shown in the first column of Fig.~\ref{fig:5}(b), where distinct kinetics- and diffusion-limited regimes are observed. At sufficiently small $\Da$, $\avg{Sh}$ scales linearly with $\Da$, approaching $\avg{Sh}\to\Da$ as $\Da\to0$ with unit slope for both geometries. This geometry-independent asymptotic behavior reflects the direct proportionality between reactive flux and the interfacial reaction-rate constant when reactions are rate-limiting. As $\Da$ increases, $\avg{Sh}$ becomes progressively less sensitive to $\Da$ and approaches diffusion-limited values of $3$ and $4$ for parallel plates and circular tubes, respectively. These values agree with our previously reported PSL results for the perfectly reactive limit \cite{Hamid2023}.

The frequency response of $\Shnd$ differs fundamentally from that of $\Nx$. Whereas $|\Nx|$ decreases with increasing frequency, $|\Shnd|$ increases because rapid forcing reduces the characteristic diffusion length near the reactive interface and produces steeper concentration gradients. For $\wstar>10$, the kinetics-limited response approaches a frequency-independent plateau because the interfacial reaction rate remains rate-limiting. In the diffusion-limited regime, by contrast, $|\Shnd|\sim\omega^{1/2}$, consistent with the classical semi-infinite Warburg response~\cite{Huang2018}. The corresponding phase approaches $\theta_{\Shnd}\to\pi/4$ non-monotonically, initially exceeding its high-frequency asymptote before returning toward $\pi/4$. This behavior reflects the transition from transport influenced by the finite pore geometry at intermediate frequencies to a locally diffusion-controlled interfacial response at high frequencies.

The $\omega^{1/2}$ growth of $|\Shnd|$ represents a fundamental departure from the steady diffusion-limited response. Under sufficiently rapid forcing, diffusion penetrates only a characteristic distance $\delta\sim\sqrt{D/\omega}$ from the interface. The corresponding mass-transfer coefficient therefore scales as $h\sim D/\delta\sim\sqrt{D\omega}$ and increases with frequency rather than approaching a constant steady-state value. Thus, the conventional diffusion-limited flux represents a steady-state limit rather than an absolute bound on transient interfacial mass transfer. Rapid transient forcing can therefore produce instantaneous interfacial fluxes that exceed the corresponding steady diffusion-limited value. We explore this implication in detail in a companion manuscript, where the consequences for transient over-limiting operation of electrochemical systems, including flow batteries, are examined~\cite{Hamid2026PRL}.

\subsection{Frequency-Dependent Flow-Induced Reactivity Bias}

Unlike $\Nx$ and $\Shnd$, the reactivity-bias vector $\Rx$ depends linearly on $\Pe$, as evidenced from Table~\ref{tab:2}. Normalization by $\Pe$ therefore isolates its dependence on $\Da$ and frequency. As shown in the first column of Fig.~\ref{fig:6}, $\avg{R}_x/\Pe$ exhibits the same two reaction regimes observed for $\avg{N}_x$. At sufficiently small $\Da$, $\avg{R}_x/\Pe$ increases linearly with $\Da$, with asymptotic slopes of $1/15$ and $1/24$ for parallel plates and circular tubes, respectively. As $\Da$ increases, the response becomes progressively less sensitive to the reaction-rate constant and approaches the diffusion-limited values $\avg{R}_x/\Pe=1/5$ and $1/6$ for parallel plates and circular tubes, respectively.

The Bode plots in the right two columns of Fig.~\ref{fig:6} show the corresponding transition in high-frequency response. In the kinetics-limited regime, $|\Rx|$ exhibits an $\omega^{-1}$ scaling characteristic of a capacitive-type response. As $\Da$ increases and the system becomes diffusion-limited, the high-frequency scaling approaches $\omega^{-1/2}$, accompanied by the corresponding Warburg-type phase response. This transition reflects a change in the rate-limiting mechanism. When interfacial kinetics are slow, the reactivity bias is controlled primarily by the reaction rate. In the diffusion-limited regime, transverse diffusion controls how rapidly the influence of the reactive interface is communicated across the pore cross-section. Consequently, $\Rx$ becomes progressively less sensitive to $\Da$ and exhibits the characteristic Warburg response.

    \section{\label{sec:Sec_05}Upscaled Transient Transport for Flow Between Non-Reactive Parallel Plates}

\begin{figure}[t!]
    \centering
    \includegraphics[width=0.9\linewidth]{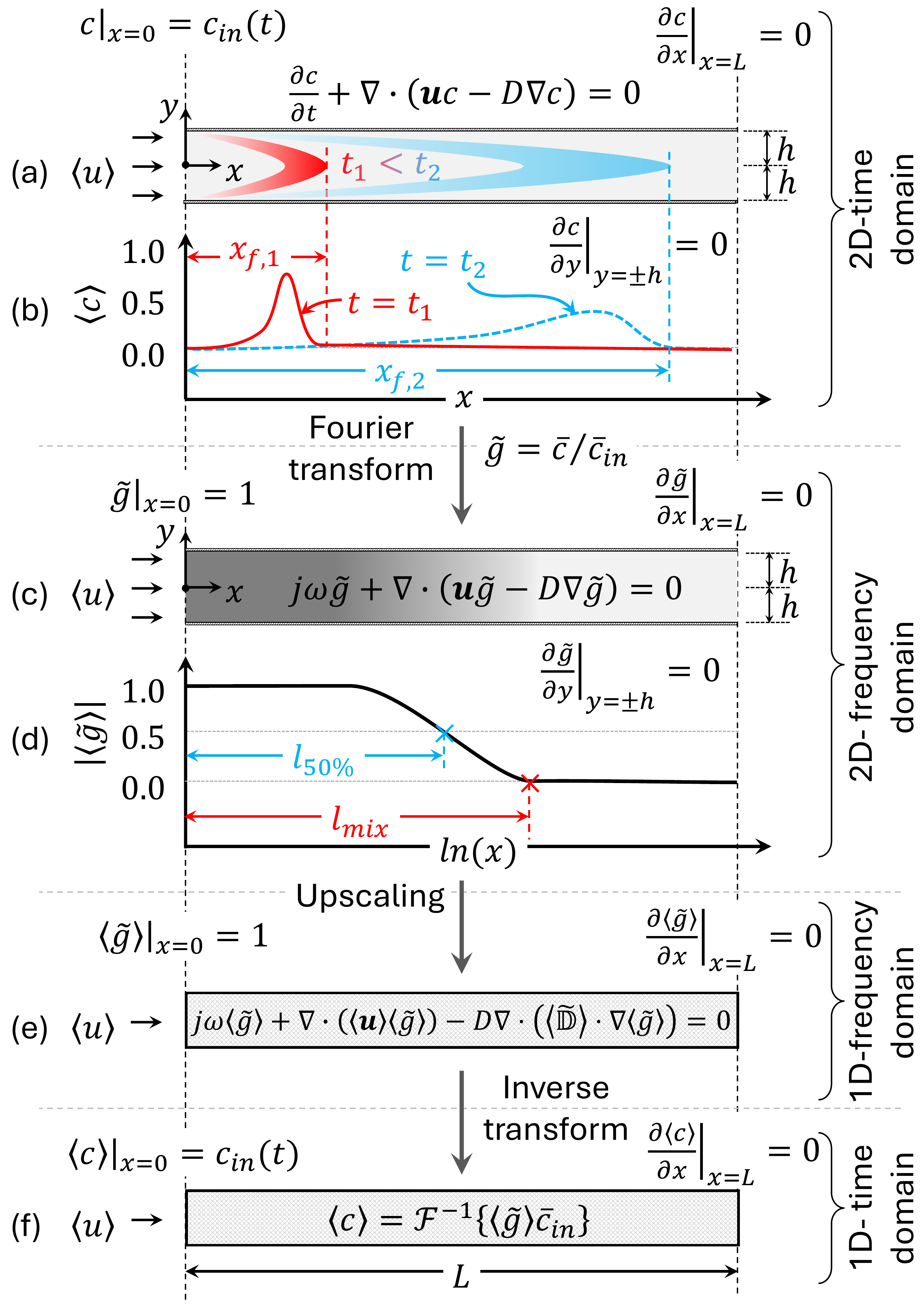}
    \caption{Schematic comparison of the pore-scale and upscaled transport models. (a) Two-dimensional (2D) Poiseuille flow of a Gaussian solute pulse between parallel plates, with concentration distributions shown at $t_1$ and $t_2>t_1$. (b) Corresponding qualitative variation of the cross-sectionally averaged concentration along the $x$ axis at $t_1$ and $t_2$, with solute-front positions $x_{f,1}$ and $x_{f,2}$. (c) Two-dimensional frequency-domain model represented by the concentration transfer function $\tf{g}=\tc/\tc_{in}$. (d) Qualitative variation of the cross-sectionally averaged transfer function $\avg{\tf{g}}$ along the $x$ axis, showing the mixing-zone length $l_{mix}$ and the location $l_{50\%}$ at which $|\avg{\tf{g}}|=0.5$. (e) One-dimensional (1D) upscaled frequency-domain model governing $\avg{\tf{g}}$ in the absence of interfacial reaction. (f) Corresponding one-dimensional upscaled time-domain response obtained through inverse Fourier transformation.}
    \label{fig:7}
\end{figure}

In this section, we investigate transient solute transport in steady, fully developed flow between two non-reactive parallel plates, corresponding to $\Da=0$. The fluid initially contains no solute, and a time-varying concentration pulse is introduced uniformly across the transverse direction at the inlet. A schematic of the system is shown in Fig.~\ref{fig:7}(a). Supporting videos (Movie 1 and Movie 2) provide visual illustrations of the corresponding pore-scale concentration evolution. We focus on the time evolution of the cross-sectional average concentration along the flow direction, or breakthrough curve, and use it to validate the omni-temporal framework against direct numerical simulations. The conventional pseudo-steady dispersion model, which assumes a time-independent dispersion coefficient, is also considered for comparison.

\begin{figure}[t!]
    \centering
    \includegraphics[width=0.85\linewidth]{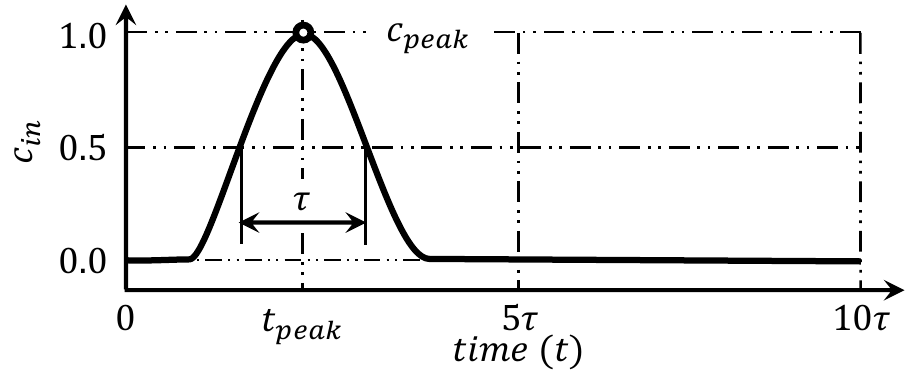}
    \caption{Gaussian inlet concentration pulse $c_{in}(t)$, with peak concentration $c_{peak}$ occurring at $t=t_{peak}$.}
    \label{fig:8}
\end{figure}

Although the proposed framework is formulated in the frequency domain and the time-domain response is subsequently recovered by inverse Fourier transformation, we first examine the breakthrough curves. Because the plates are infinitely extended in the $z$-direction, the pore-scale problem reduces to two dimensions (Fig.~\ref{fig:7}(a,c)). The framework permits arbitrary inlet concentration histories, but a Gaussian pulse is used here because its smooth temporal variation is advantageous for numerical Fourier transformation. Accordingly, the inlet concentration is prescribed as
\begin{align}
    c_{in}(t) & = c_{peak}\exp\left[-\ln(2)\left(\frac{t-t_{peak}}{\tau/2}\right)^2\right] \label{eq:51}    
\end{align}
where $c_{peak}$, $t_{peak}$, and $\tau$ denote the peak concentration, peak time, and full-width at half-maximum (FWHM) of the pulse, respectively. The inlet pulse defined by Eq.~\eqref{eq:51} is shown schematically in Fig.~\ref{fig:8}.

The breakthrough curves are obtained using the three approaches described below. The same parallel-plate system is then examined in the frequency domain to characterize its transient transport response.
\begin{itemize}[leftmargin=*]
    \item \textit{Two-Dimensional Direct Numerical Simulation} (2D-DNS): The two-dimensional form of Eq.~\eqref{eq:2} is solved directly in the time domain to obtain $c(x,y,t)$, with the governing equation and boundary conditions illustrated schematically in Fig.~\ref{fig:7}(a). The corresponding frequency-domain problem is obtained from Eq.~\eqref{eq:4} using the transfer function $\tf{g}(x,y,\omega)=\tc(x,y,\omega)/\tc_{in}(\omega)$, as illustrated in Fig.~\ref{fig:7}(c). Once $c(x,y,t)$ and $\tf{g}(x,y,\omega)$ are obtained, the cross-sectional averages are calculated as $\avg{c}(x,t)=(1/2h)\int^{+h}_{-h}c(x,y,t)\,dy$ and $\avg{\tf{g}}(x,\omega)=(1/2h)\int^{+h}_{-h}\tf{g}(x,y,\omega)\,dy$.
    
    \item \textit{One-Dimensional Omni-Temporal Dispersion} (1D-OTD): The one-dimensional frequency-domain upscaled equation is solved for the cross-sectionally averaged transfer function $\avg{\tf{g}}(x,\omega)=\avg{\tc}(x,\omega)/\tc_{in}(\omega)$. The corresponding governing equation and boundary conditions are illustrated schematically in Fig.~\ref{fig:7}(e). The frequency-dependent dispersion coefficient $\Dxx$ listed in Table~\ref{tab:2} is used to describe the effective axial dispersion. The corresponding time-domain solution is then recovered through inverse Fourier transformation (iFFT), as shown in Fig.~\ref{fig:7}(f).
    
    \item \textit{One-Dimensional Pseudo-Steady Dispersion} (1D-PSD): This approach is identical to the 1D-OTD formulation except that the dispersion coefficient is assumed to be independent of time and frequency and is prescribed using the PSL expression for $\avg{\Dbb}_{xx}$. The 1D-PSD formulation therefore represents the conventional pseudo-steady, or quasi-steady, treatment of hydrodynamic dispersion.
\end{itemize}
Figure~\ref{fig:7} summarizes the governing equations and boundary conditions for the pore-scale and upscaled formulations in both the time and frequency domains. The reduced governing equations and boundary conditions for each approach are provided in Sec.~H of the SI, while details of the numerical solution procedures and convergence tests are provided in Sec.~J of the SI.

\subsection{Time-Domain Results}

\begin{figure}[t!]
    \centering
    \includegraphics[width=0.9\linewidth]{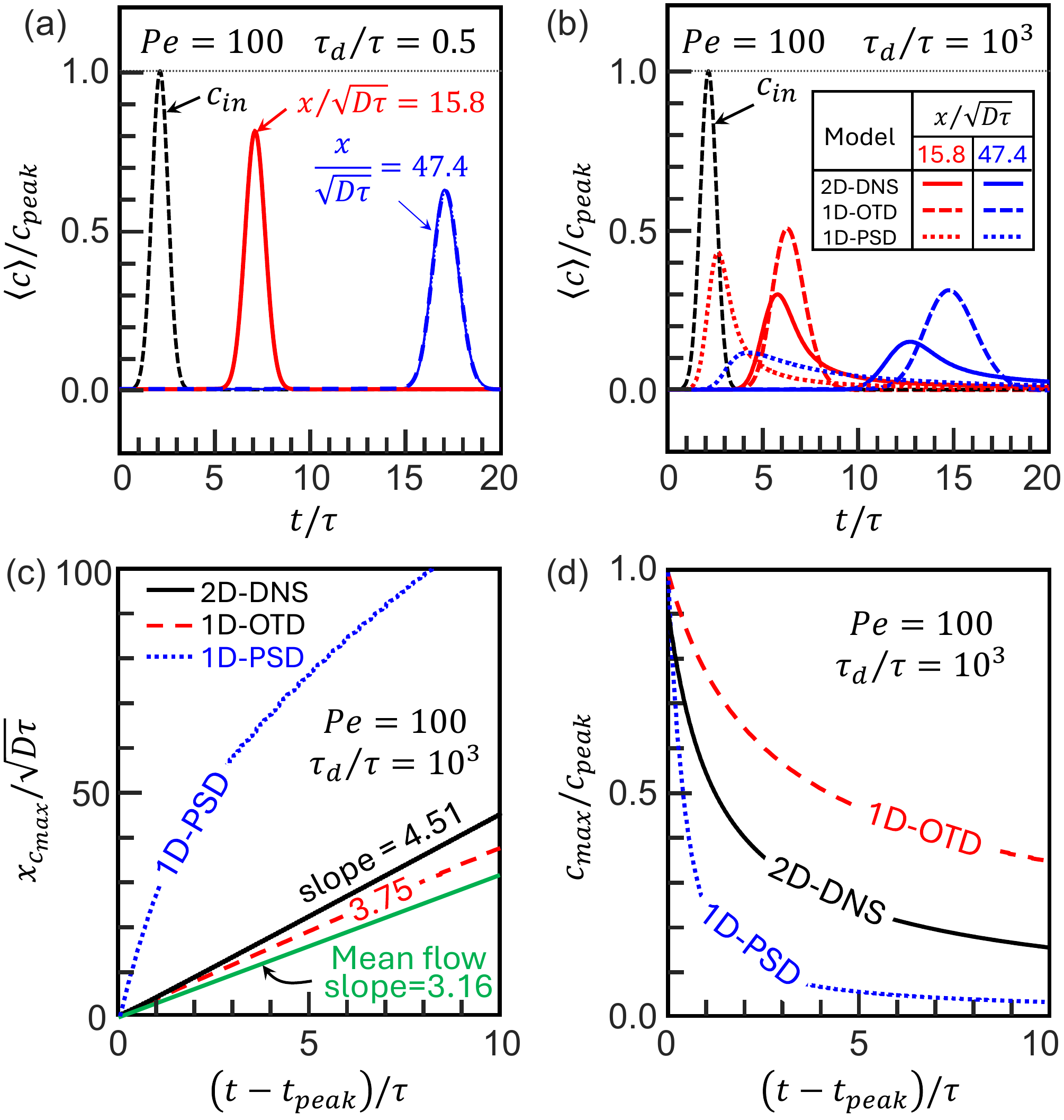}
    \caption{Time-domain response to Gaussian inlet pulses for $\Pe=100$. Variation of the normalized average concentration $\avg{c}/c_{peak}$ with $t/\tau$ for (a) $\tau_d/\tau=0.5$ and (b) $\tau_d/\tau=10^3$ at $x/\sqrt{D\tau}=15.8$ (red) and $x/\sqrt{D\tau}=47.4$ (blue). Results from 2D-DNS, 1D-OTD, and 1D-PSD are shown by solid, dashed, and dotted lines, respectively. (c) Variation of the nondimensional peak location $x_{c_{\max}}/\sqrt{D\tau}$ with time for $\tau_d/\tau=10^3$. (d) Corresponding variation of the normalized peak concentration $c_{\max}/c_{peak}$. In (c) and (d), the time axis is shifted such that $t=t_{peak}$ corresponds to the origin.}
    \label{fig:9}
\end{figure}

\begin{figure}[t!]
    \centering
    \includegraphics[width=0.9\linewidth]{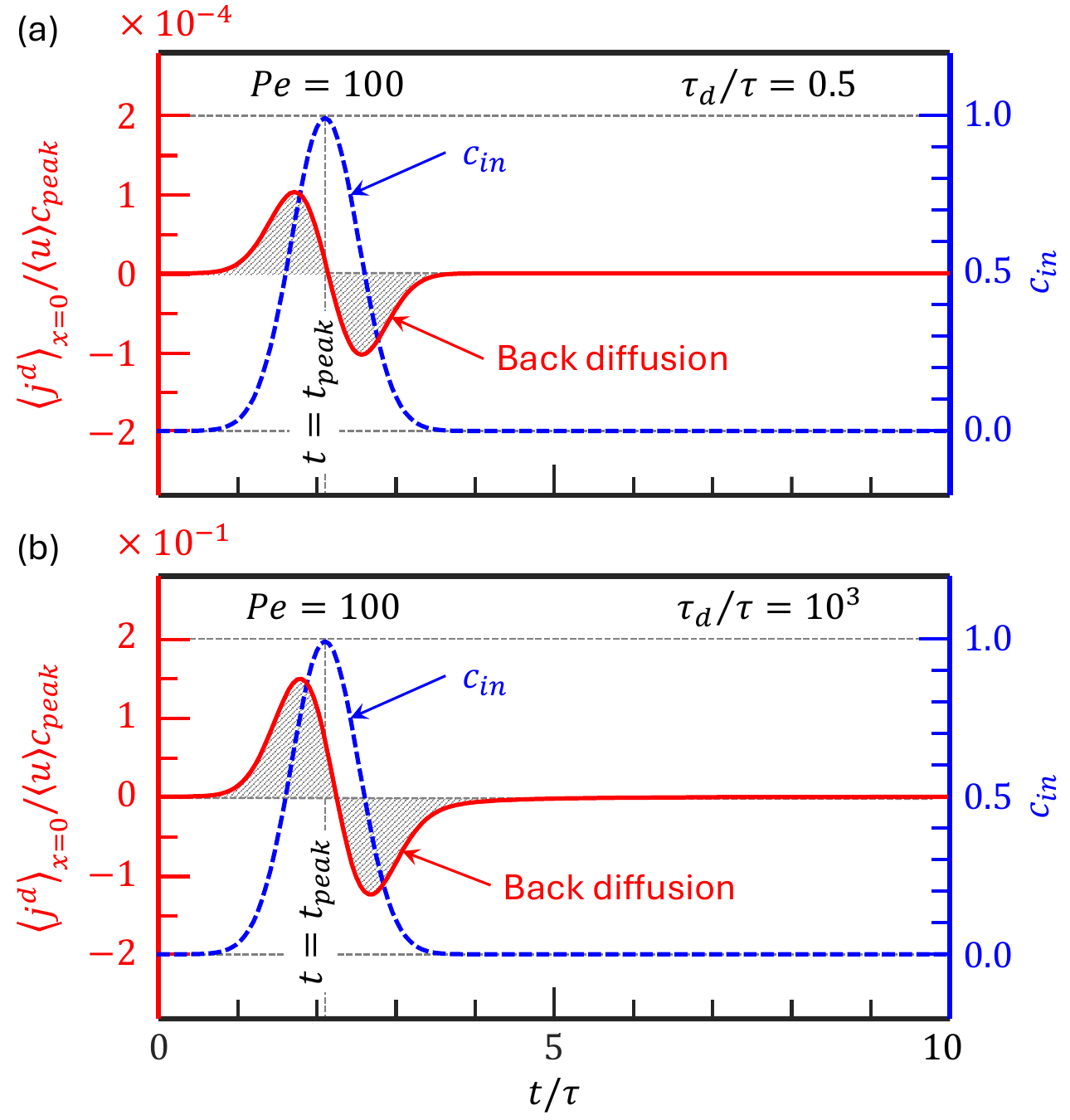}
    \caption{Time variation of the cross-sectionally averaged $x$-directed diffusive flux at the inlet ($x=0$) for $\Pe=100$ with (a) $\tau_d/\tau=0.5$ and (b) $\tau_d/\tau=10^3$.}
    \label{fig:10}
\end{figure}

We first compare the time-domain response predicted by the omni-temporal upscaled model (1D-OTD) with the corresponding 2D direct numerical simulation (2D-DNS). Two cases at $\Pe=100$ are considered: a long-duration pulse with $\tau_d/\tau=0.5$, for which the pulse duration is comparable to the pore-scale diffusion time $\tau_d=h^2/D$, and a short-duration pulse with $\tau_d/\tau=10^3$, for which the pulse duration is much shorter than $\tau_d$. Supporting videos in the SI illustrate the evolution of the pore-scale concentration field from 2D-DNS.

For the long-duration pulse (see \textit{Movie 1}), solute propagates approximately at the mean flow velocity and spreads symmetrically about the pulse peak, as shown in Fig.~\ref{fig:9}(a). This behavior is consistent with classical Taylor dispersion~\cite{Taylor1953} in the long-time limit, when transverse diffusion has sufficient time to redistribute solute across streamlines~\cite{Han1985}. In contrast, the short-duration pulse initially develops a parabolic concentration profile under the nonuniform velocity field (see \textit{Movie 2}). Transverse diffusion subsequently smooths this profile, but the solute remains localized near the inlet for a substantial period and leaves a pronounced long tail after the inlet concentration has returned to zero. Similar long-tail behavior has been reported previously~\cite{Porta2013,Gist1990,Bacri1990}.

Figures~\ref{fig:9}(a,b) compare the average concentration $\avg{c}$ at two axial positions, $x/\sqrt{D\tau}=15.8$ and $x/\sqrt{D\tau}=47.4$, obtained from 2D-DNS, 1D-OTD, and 1D-PSD. For $\tau_d/\tau=0.5$, all three models give nearly identical results, with approximately symmetric pulse spreading and downstream propagation close to the mean flow velocity. For $\tau_d/\tau=10^3$, 2D-DNS instead predicts strongly asymmetric spreading, whereas both upscaled models retain more symmetric profiles. The pseudo-steady model shows the largest discrepancy, significantly underpredicting the travel time of the pulse peak to a given downstream location. The omni-temporal model predicts the propagation more accurately, demonstrating the importance of frequency-dependent dispersion. Nevertheless, 1D-OTD still mildly overpredicts both the peak concentration and the arrival time relative to 2D-DNS.

Figure~\ref{fig:9}(c) further shows that 2D-DNS and 1D-OTD predict an approximately linear relation between the peak position $x_{c_{\max}}/\sqrt{D\tau}$ and nondimensional time $t/\tau$, corresponding to constant propagation velocities of $4.51$ and $3.75$, respectively. Thus, the omni-temporal model underpredicts the propagation velocity by approximately $15\%$. In contrast, the pseudo-steady model predicts a nonlinear trajectory, indicating progressive deceleration of the pulse. Its initial nondimensional propagation velocity is approximately $35$, exceeding the DNS value by more than $650\%$. Figure~\ref{fig:9}(d) further shows that the pseudo-steady model systematically underpredicts the average concentration because of excessive axial spreading. The omni-temporal model, in contrast, overpredicts the peak concentration, primarily because of two effects: (i) the long-tail concentration distribution, which produces back diffusion at the inlet and is not accounted for in the 1D-OTD model; and (ii) entry-region effects associated with imposing a spatially uniform inlet concentration at $x=0$. The long-tail effect is examined below in the time domain, while the entry-region effect is examined later in this section using frequency-domain analysis.

The long concentration tail produces a nonzero concentration gradient at the inlet after pulse injection and therefore generates a diffusive flux opposite to the bulk flow. We refer to this effect as \textit{back-diffusion}~\cite{Warden2025}. The inlet-averaged back-diffusion flux obtained from the 2D-DNS simulations is
\begin{equation}
    \avg{j^d}_{x=0} = -\frac{1}{2h}\int_{-h}^{+h} D\left.\left(\frac{dc}{dx}\right)\right|_{x=0}dy \label{eq:52}
\end{equation}
To quantify its magnitude, $\avg{j^d}_{x=0}$ is normalized by the peak advective flux $\avg{u}c_{peak}$. Figure~\ref{fig:10} shows that for $\tau_d/\tau=0.5$, the back-diffusion flux remains negligible, contributing less than $0.01\%$ of the peak advective flux over the simulated interval $0\le t\le10\tau$. For $\tau_d/\tau=10^3$, substantial back diffusion, exceeding $13\%$ of the peak advective flux, persists for approximately $2\tau$ beginning near $t\approx t_{peak}$. This back diffusion reduces the amount of solute retained between the plates and decreases the effective downstream propagation rate. Models that neglect this effect therefore tend to overpredict peak concentration and downstream propagation velocity. Accounting for back diffusion is an important direction for improving the omni-temporal framework.

\subsection{Frequency-Domain Results}

\begin{figure}[t!]
    \centering
    \includegraphics[width=0.85\linewidth]{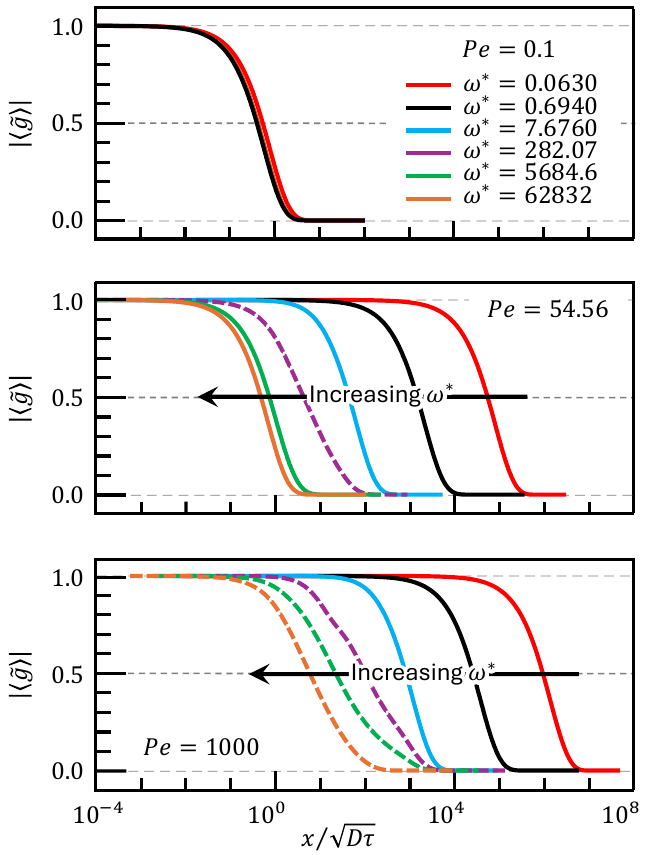}
    \caption{Magnitude of the cross-sectionally averaged concentration transfer function $|\avg{\tf{g}}|$ obtained from 2D-DNS as a function of longitudinal distance for Poiseuille flow between non-reactive parallel plates at different $\Pe$ and $\wstar$.}
    \label{fig:11}
\end{figure}

\begin{figure}[t!]
    \centering
    \includegraphics[width=0.8\linewidth]{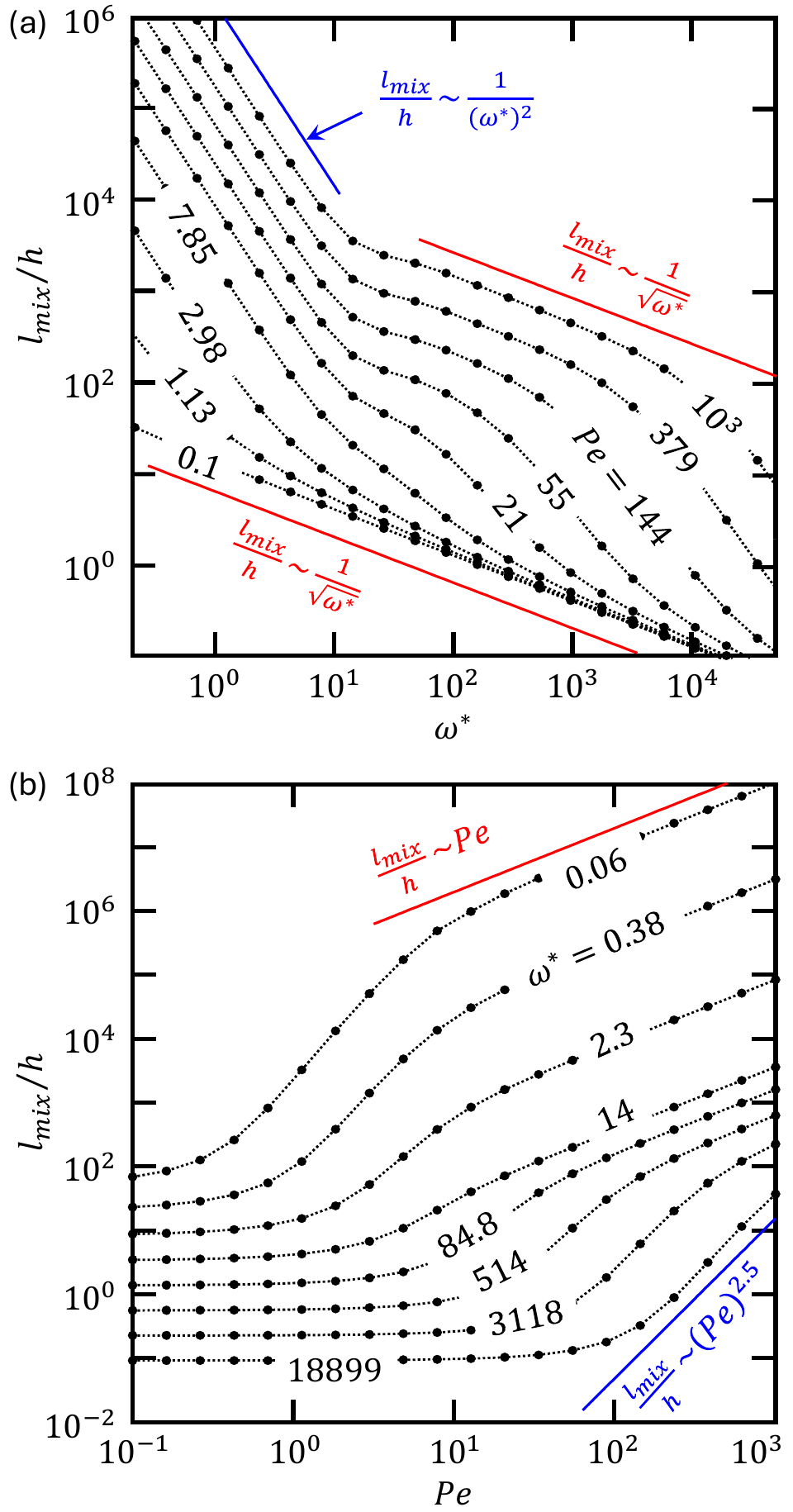}
    \caption{Variation of the nondimensional mixing length $l_{mix}/h$ (a) with $\wstar$ at different $\Pe$ and (b) with $\Pe$ at different $\wstar$ for Poiseuille flow between non-reactive parallel plates.}
    \label{fig:12}
\end{figure}

\begin{figure*}[t!]
    \centering
    \includegraphics[width=0.85\linewidth]{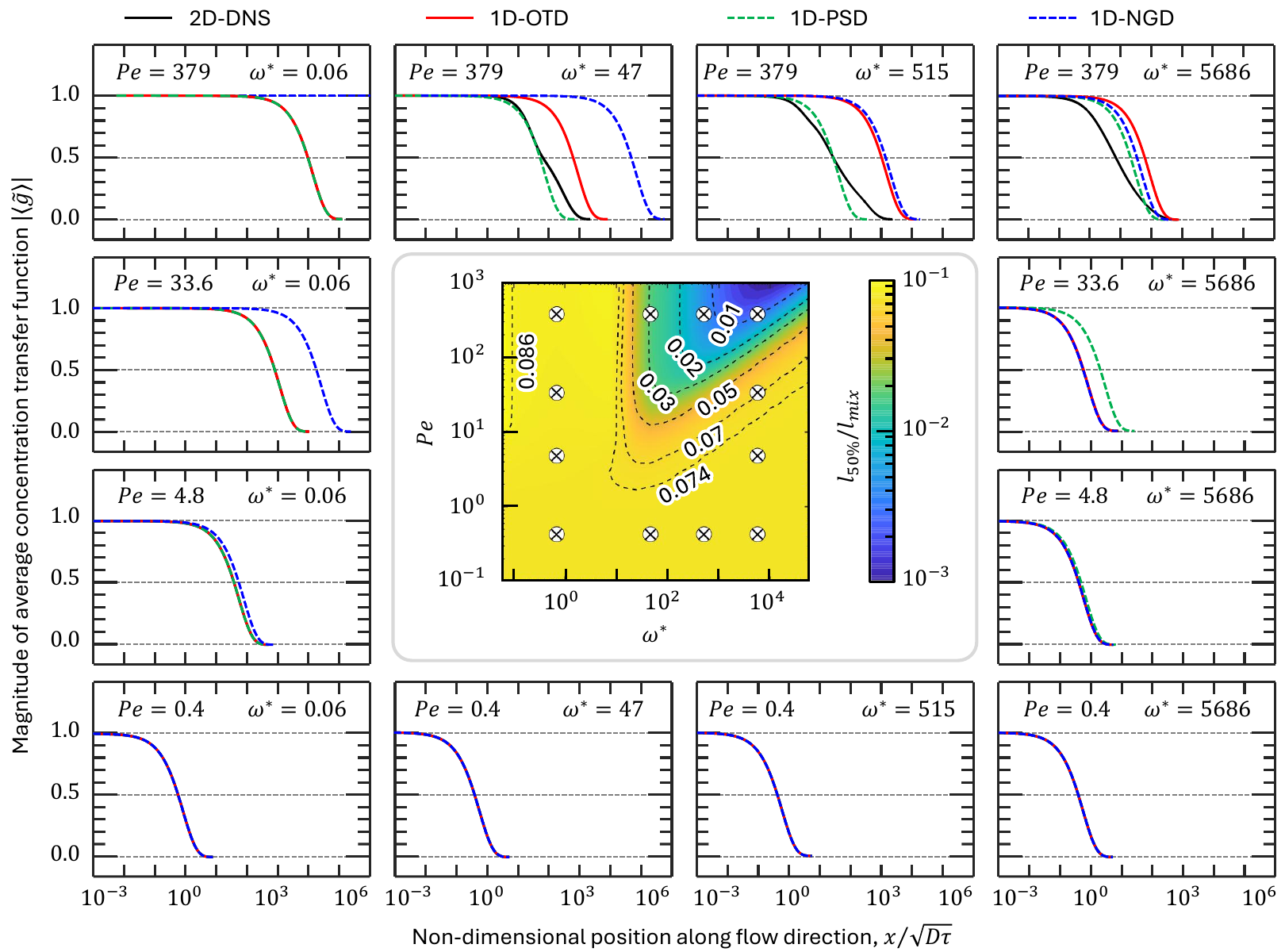}
    \caption{Frequency-domain comparison of the pore-scale and upscaled transport models. The center panel shows a contour plot of $l_{50\%}/l_{mix}$ as a function of $\Pe$ and $\wstar$. The $\otimes$ symbols identify the combinations of $\Pe$ and $\wstar$ shown in the surrounding panels. These panels compare the spatial variation of $|\avg{\tf{g}}|$ predicted by 2D-DNS (black solid), 1D-OTD (red solid), 1D-PSD (green dashed), and 1D-NGD (blue dashed).}
    \label{fig:13}
\end{figure*}

\begin{figure}[t!]
    \centering
    \includegraphics[width=0.85\linewidth]{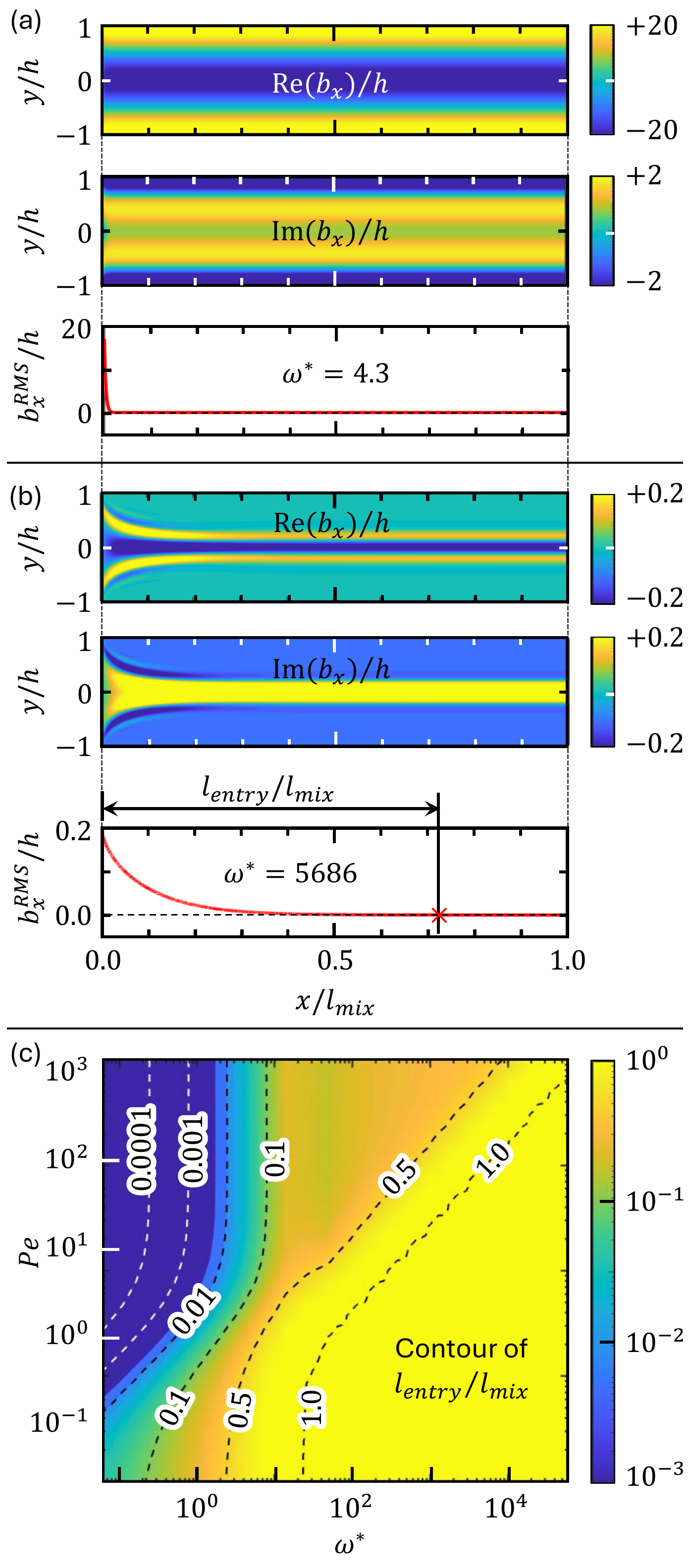}
    \caption{Entry-region behavior of the closure field $b_x$. Contours show the real and imaginary components of $b_x$, together with the axial variation of $b_x^{RMS}$, for $\Pe=379$ at (a) $\wstar=4.3$ and (b) $\wstar=5686$. (c) Contour plot of the normalized entry length $l_{entry}/l_{mix}$ as a function of $\Pe$ and $\wstar$.}
    \label{fig:14}
\end{figure}

We next examine the frequency-domain response predicted by the omni-temporal upscaled model (1D-OTD) and compare it with high-fidelity pore-scale simulations (2D-DNS). We investigate 480 combinations of $\Pe$ and $\omega^*$. For comparison, we also consider two simplified one-dimensional models. The first is the conventional pseudo-steady model (1D-PSD). The second neglects hydrodynamic dispersion and is therefore referred to as the neglected-dispersion (1D-NGD) model. The 1D-NGD model has the same governing-equation structure as the 1D-OTD and 1D-PSD models but uses the molecular diffusivity as the effective dispersion coefficient, i.e., $\avg{\Dbb}_{xx}=1$. The frequency-domain results are expressed through the transfer function $\avg{\tf{g}}(x,\omega)$, defined as the ratio of the Fourier-transformed cross-sectional average concentration to the Fourier-transformed inlet concentration,
$\avg{\tf{g}}(x,\omega)=\avg{\tc}(x,\omega)/\tc_{in}(\omega)$. Equivalently, $\avg{\tf{g}}$ is the frequency response to a unit impulse at the inlet under the corresponding boundary conditions. For the upscaled models, $\avg{\tf{g}}$ is obtained directly from the governing equations. For 2D-DNS, it is obtained by averaging the two-dimensional transfer function $\tf{g}$ across the transverse direction at a given axial location, $\avg{\tf{g}}=(1/2h)\int_{-h}^{+h}\tf{g}\,dy$.

Figure~\ref{fig:11} shows the magnitude of $\avg{\tf{g}}$ obtained from 2D-DNS for different $Pe$ and nondimensional frequencies $\omega^*$. The axial position is normalized by the characteristic diffusion length $\sqrt{D\tau}$, where $\tau=2\pi/\omega$ is the oscillation period associated with $\omega$. Two characteristic length scales can be identified for a given frequency: the median propagation length $l_{50\%}$, defined by $|\avg{\tf{g}}|=0.5$, and the mixing-zone length $l_{mix}$, beyond which $|\avg{\tf{g}}|$ becomes negligibly small (see Fig.~\ref{fig:7}(d)). In these calculations, $|\avg{\tf{g}}|=10^{-6}$ was used to define $l_{mix}$ (this threshold is arbitrarily selected). These length scales provide useful measures of signal propagation and attenuation. To limit attenuation of the inlet signal, the flow-path length $L$ should remain small relative to $l_{50\%}$. Conversely, achieving substantial attenuation or mixing requires $L>l_{mix}$. When $\Pe<1$, the normalized response in Fig.~\ref{fig:11} depends weakly on frequency because the relevant diffusion length is $\sqrt{D\tau}$; after normalization by this length, the response approaches a common diffusion-dominated form. For $\Pe>1$, advection increasingly extends the downstream response through shear-assisted transport. Under these conditions, $l_{50\%}$ and $l_{mix}$ exhibit an approximately fixed ratio, except for the cases shown by dashed curves in Fig.~\ref{fig:11}. The center panel of Fig.~\ref{fig:13} shows that most cases satisfy $l_{50\%}\approx0.074\,l_{mix}$. At simultaneously large $\Pe$ and $\omega^*$, however, this ratio decreases significantly. This reduction is associated with back diffusion from the long-tail concentration distribution identified from the time-domain results.

Figure~\ref{fig:12}(a) shows that for $\Pe<1$, the mixing length scales as $1/\sqrt{\omega^*}$, consistent with diffusion-dominated transport because $\sqrt{D\tau}\sim\sqrt{2\pi D/\omega^*}$. Similar scaling is observed for $\Pe\gtrsim100$ over an intermediate frequency range. At lower frequencies, $\omega^*\lesssim10$, the mixing length instead follows $l_{mix}\sim(\omega^*)^{-2}$. Figure~\ref{fig:12}(b) further shows that $l_{mix}/h$ varies approximately linearly with $\Pe$ at low frequencies for $\Pe\gtrsim10$. At high $\Pe$ and $\omega^*\to\infty$, the scaling becomes super-linear, with $l_{mix}/h\sim\Pe^{2.5}$.

The spatial variation of $|\avg{\tf{g}}|$ for different $\Pe$ and $\omega^*$ is shown in the surrounding panels of Fig.~\ref{fig:13}. Each panel compares 2D-DNS with the upscaled models: 1D-OTD, 1D-PSD, and 1D-NGD. At low P\'eclet number ($\Pe\ll1$), hydrodynamic dispersion is negligible, causing all four models to collapse onto the same solution, nearly independently of $\omega^*$. In the low-frequency limit ($\omega^*\ll1$, left panels), both 1D-OTD and 1D-PSD closely reproduce the 2D-DNS results, whereas 1D-NGD increasingly overpredicts $l_{mix}$ as $\Pe$ increases. This agreement reflects the validity of the long-time pseudo-steady limit at low frequency. As $\Pe$ increases, hydrodynamic dispersion becomes increasingly important, and neglecting it in 1D-NGD produces substantial errors. In the high-frequency limit ($\omega^*\gg1$, right panels), the pseudo-steady approximation underlying 1D-PSD breaks down, leading to increasing overprediction of $l_{mix}$ with $\Pe$. In contrast, 1D-OTD remains accurate because it incorporates the frequency dependence of the dispersion coefficient. Under strongly advective conditions ($\Pe\gg100$), all upscaled models deviate from 2D-DNS, particularly at large $\omega^*$. These discrepancies arise primarily from the long-tail concentration distributions and the associated back diffusion.

Finally, we examine the entry-region effects associated with the inlet boundary condition $\left.\tf{g}\right|_{x=0}=1$ used in the pore-scale transport problem. From the 2D-DNS simulations, a spatially varying closure field $b_x(x,y,\omega)$ is calculated as
\begin{equation}
    b_x(x,y,\omega) = \frac{\tc'}{\partial\avg{\tc}/\partial x} = \frac{\tf{g}-\avg{\tf{g}}}{\partial\avg{\tf{g}}/\partial x} \label{eq:53}
\end{equation}
In the absence of entry effects, $b_x(x,y,\omega)$ should remain invariant along the streamwise direction for the present geometry, consistent with the periodicity assumed in the 1D-OTD model. The entry region is therefore defined as the portion of the domain over which $b_x$ varies significantly with $x$. To quantify its extent, the root-mean-square (RMS) value of $b_x$ is evaluated as
\begin{equation}
    b_x^{RMS} = \sqrt{\frac{1}{2h}\int_{-h}^{+h}\left(b_x-\left.b_x\right|_{x\to\infty}\right)^2dy} \label{eq:54}
\end{equation}
Here, $\left.b_x\right|_{x\to\infty}$ denotes the fully developed value of $b_x$, which is independent of $x$.

Figures~\ref{fig:14}(a,b) show the spatial variation of the real and imaginary components of $b_x(x,y,\omega)$ for $\Pe=379$ at two frequencies. At low frequency ($\omega^*=4.3$), $b_x$ exhibits an approximately invariant transverse profile over most of the mixing length. At high frequency ($\omega^*=5686$), strong transverse gradients develop near the inlet and evolve within a thin boundary layer before approaching an invariant downstream profile. This behavior is analogous to that of a classical convective boundary layer~\cite{Prandtl1938}. To quantify the entry effects, $b_x^{RMS}$ was evaluated for $\Pe=379$ at $\omega^*=4.3$ and $\omega^*=5686$. At low frequency ($\omega^*=4.3$), the entry region occupies less than $1\%$ of the mixing length, whereas at $\omega^*=5686$ it extends over more than $70\%$. Figure~\ref{fig:14}(c) further shows that the normalized entry length $l_{entry}/l_{mix}$ is generally largest at high frequency and low $\Pe$. Figure~\ref{fig:13} nevertheless indicates that the upscaled models remain relatively accurate in this regime because the flow-induced dispersive contribution is small. The largest discrepancies instead occur at simultaneously high $\Pe$ and $\omega^*$, where moderate entry-region effects combine with the long-tail concentration distributions and associated back diffusion.

    \section{\label{sec:Sec_06}Transient Transport in a Model Porous Medium}

\begin{figure}[t!]
    \centering
    \includegraphics[width=0.9\linewidth]{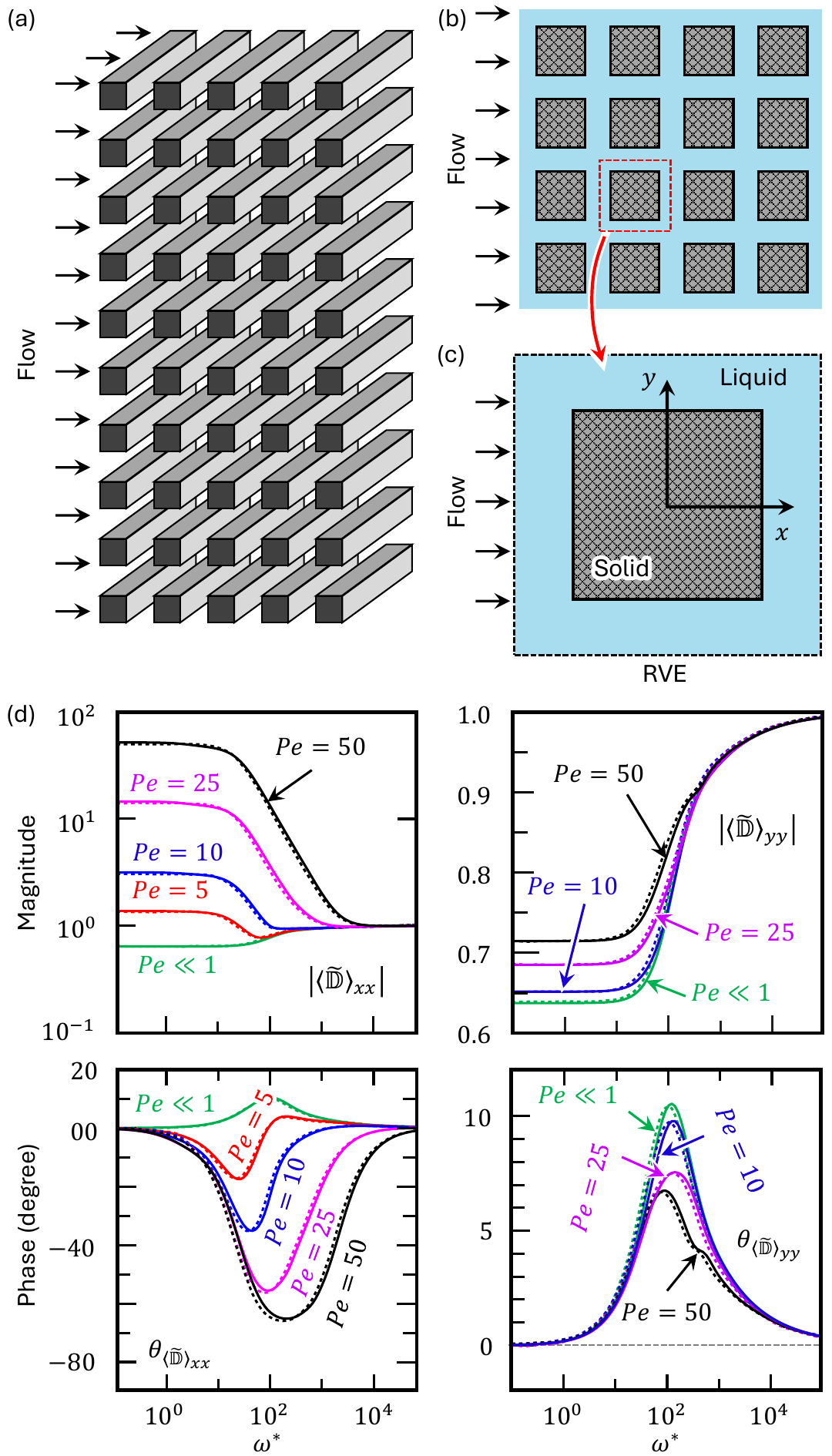}
    \caption{(a) Three-dimensional representation of flow through a periodic array of square rods, with the rods assumed to extend infinitely in the axial direction. (b) Corresponding two-dimensional representation of the square-rod array under cross flow. (c) RVE used for the upscaling analysis. (d) Frequency response of the dispersion coefficients in the non-reactive limit ($\Da=0$) at $\eps=0.5$. The top and bottom rows show the magnitude and phase, respectively, while the left and right columns correspond to the longitudinal $\Dxx$ and transverse $\Dyy$ components. Results from the present omni-temporal framework are shown as solid lines, while those of Parada and Ramirez~\cite{ValdesParada2011} are shown as dashed lines.}
    \label{fig:15}
\end{figure}

\begin{figure}[t!]
    \centering
    \includegraphics[width=0.9\linewidth]{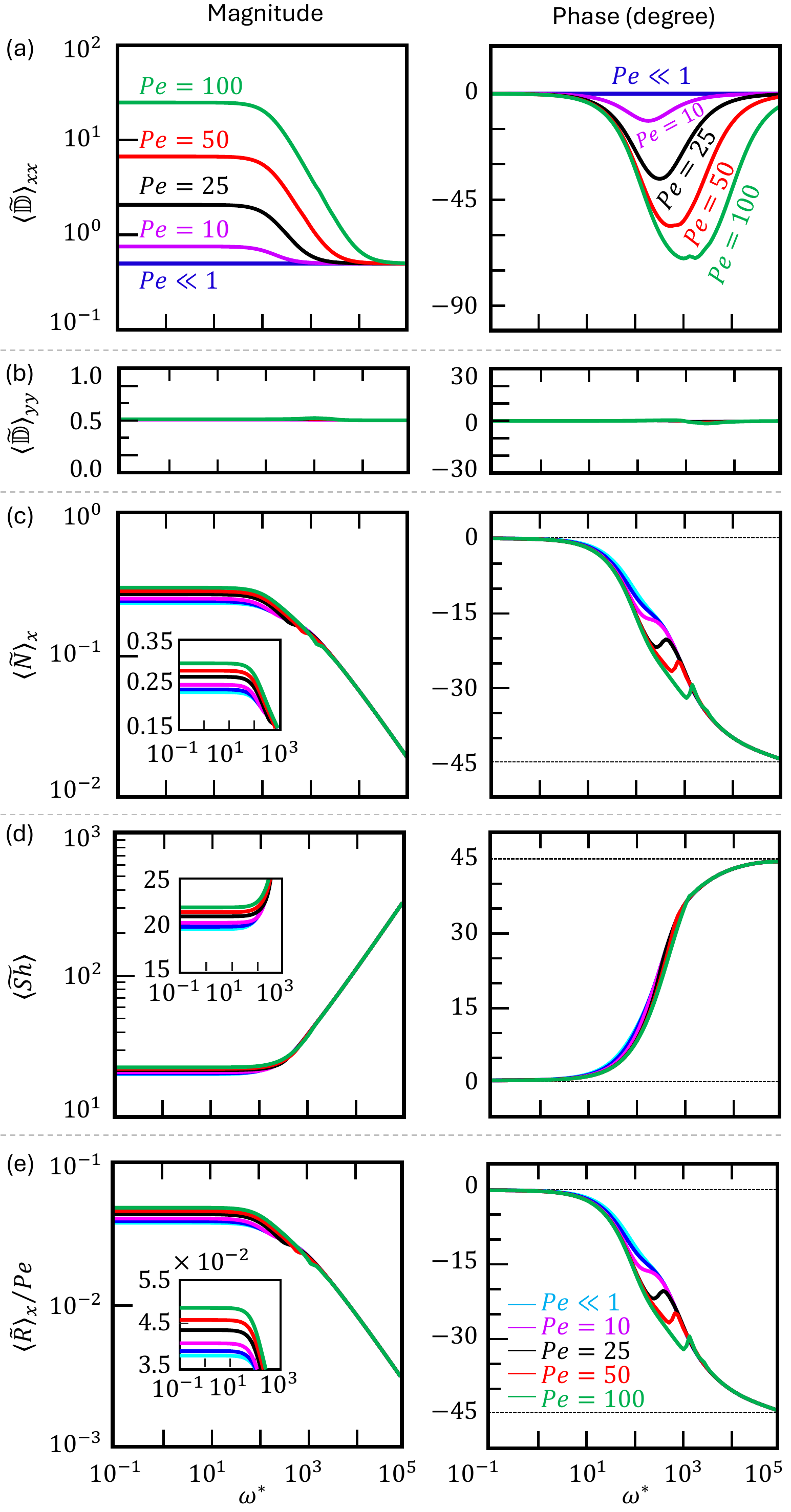}
    \caption{Frequency response of the upscaled transport coefficients in the perfectly reactive limit ($\Da\to\infty$) for a periodic array of square rods under cross flow at $\eps=0.5$. (a) Dispersion coefficients: the left and middle panels show the magnitude and phase of the longitudinal component $\Dxx$, respectively, while the right panels show the magnitude (top) and phase (bottom) of the transverse component $\Dyy$. (b) Additional upscaled transport coefficients: advection-suppression transfer function $\Nx$ (left), spectral Sherwood number $\Shnd$ (middle), and reactivity-bias vector $\Rx$ (right). The top and bottom rows show the magnitude and phase, respectively. Insets in the magnitude plots show enlarged views of the pseudo-steady limit.}
    \label{fig:16}
\end{figure}

We now apply the omni-temporal framework to transient solute transport in a model porous medium with a more complex pore geometry. The medium consists of a periodic array of square rods under cross flow with porosity $\eps=0.5$, as illustrated in Fig.~\ref{fig:15}(a). Because the rods extend infinitely in the axial direction, the system is represented by the two-dimensional cross-section shown in Fig.~\ref{fig:15}(b), with the corresponding representative volume element (RVE) shown in Fig.~\ref{fig:15}(c). Unlike the parallel-plate and circular-tube geometries considered in Sec.~\ref{sec:Sec_04}, the closure problems do not admit simple analytical solutions and are solved numerically using the finite-volume method with central differencing. The frequency-dependent transport coefficients are obtained by solving the closure problems summarized in Table~\ref{tab:1} and evaluating Eqs.~\eqref{eq:43}--\eqref{eq:46}.

For the square-rod arrays considered here, the characteristic length scale used to define the nondimensional parameters is the equivalent spherical diameter, $\lP=6\Vs/A_s$, where $\Vs$ and $A_s$ denote the solid volume and solid--liquid interfacial area of one rod per unit axial length, respectively. Following Parada and Ramirez~\cite{ValdesParada2011}, the P\'eclet number is defined as
\begin{equation}
   \Pe = \dfrac{\avg{u}\lP}{D}\left(\frac{\eps}{1-\eps}\right). \label{eq:55}
\end{equation}
The factor $\eps/(1-\eps)$ is unity for $\eps=0.5$ and does not change the reported values of $\Pe$. This factor is included solely to enable direct correspondence with the results of Parada and Ramirez~\cite{ValdesParada2011}.

The steady, incompressible, inertia-free pore-scale velocity field $\uu$ is obtained from the two-dimensional Stokes equations in stream-function form, with $\nabla^4\psi=0$, impermeability and no-slip conditions on the rod surfaces, and periodic conditions at the unit-cell boundaries. The periodic stream function includes a constant jump corresponding to the imposed mean flow. This calculated velocity field is then used in the advection term of the frequency-dependent closure problems. For each $\wstar>0$, the closure variables $a$ and $\bcv$ are solved with periodic boundary conditions at the unit-cell boundaries and the Robin condition on the rod surfaces, using $k=0$ for non-reactive rods and $k\to\infty$ for perfectly reactive rods. Details of the discretization and numerical implementation used are provided in our earlier work~\cite{Hamid2023}.

\subsection{Dispersion in a Non-Reactive Porous Medium}

In the non-reactive limit, the reaction-related transfer functions vanish identically, $\Nnd=\Shnd=\Rnd=0$. Specifically, $a(\xl,\omega)=0$ when $\Da=0$, so $\Nnd$ and $\Shnd$ vanish (Sec.~D in the SI); moreover, $\bcv$ satisfies a homogeneous Neumann condition, $\nhat\cdot\nabla\bcv=0$, so that the surface integral defining $\Rnd$ also vanishes. Consequently, only the dispersion tensor remains nonzero, and Fig.~\ref{fig:15}(d) reports its longitudinal and transverse components, $\Dxx$ and $\Dyy$.

The frequency response shown in Fig.~\ref{fig:15}(d) reveals distinct longitudinal and transverse dispersion behavior for flow through the periodic square-rod array in the non-reactive limit ($\Da=0$). In the low-frequency limit ($\wstar\to0$), both components approach their pseudo-steady values. In this limit, solute has sufficient time to sample the pore-scale velocity field and diffuse across streamlines, producing dispersion coefficients that depend strongly on $\Pe$. The close agreement with the results of Parada and Ramirez~\cite{ValdesParada2011} confirms that the present formulation recovers the established asymptotic behavior.

As frequency increases, the values of dispersion coefficients progressively depart from their pseudo-steady values because the imposed time scale becomes comparable to, or shorter than, the pore-scale diffusion time. The concentration field therefore has insufficient time to fully adjust to the changing macroscale forcing. Consequently, the longitudinal dispersion coefficient decreases with increasing frequency, reflecting the reduced ability of transverse diffusion to couple pore-scale velocity variations to axial spreading. This behavior further demonstrates the diffusion-mediated nature of hydrodynamic dispersion observed for the canonical geometries in Sec.~\ref{sec:Sec_04}.

The transverse dispersion coefficient undergoes a similar transition, although its frequency dependence differs because of the pore geometry and flow orientation. At low $\Pe$, transverse transport is governed primarily by molecular diffusion and depends only weakly on flow. As $\Pe$ increases, advection interacts more strongly with pore geometry, producing increasingly distinct longitudinal and transverse responses. In the low-frequency limit, both phases approach zero, consistent with pseudo-steady transport. At higher frequencies, the phases depart from zero because pore-scale diffusion requires finite time to redistribute solute, providing a signature of transient transport in the pore space.

\subsection{Dispersion and Reaction in a Perfectly Reactive Porous Medium}

In the perfectly reactive limit, the symmetry of the square-rod array forces the off-diagonal dispersion components and the transverse components of the vectorial coefficients to vanish, $\Dxy=\Dyx=\Ny=\Ryc=0$ (see Section~K of the SI). The remaining coefficients that can be nonzero are therefore the longitudinal and transverse dispersion components, $\Dxx$ and $\Dyy$, the longitudinal components of the advection-suppression and reactivity-bias vectors, $\Nx$ and $\Rx$, and the scalar spectral Sherwood number $\Shnd$.

Figure~\ref{fig:16} shows the frequency-dependent transport coefficients for the periodic square-rod array in the perfectly reactive limit ($\Da\to\infty$). Interfacial reaction modifies the pore-scale concentration field and couples advection, diffusion, and reaction. As in the non-reactive case, the dispersion coefficients approach their pseudo-steady values as $\wstar\to0$. Relative to the non-reactive limit, interfacial reaction suppresses both $\Dxx$ and $\Dyy$ by altering the pore-scale concentration distribution. In this limit, the gradient-driven closure field $\bcv$ approaches zero at the reactive interface, causing the tortuosity contribution in Eq.~\eqref{eq:47} to vanish. The remaining dispersion response is therefore determined by molecular diffusion and hydrodynamic dispersion.

The frequency dependence of the additional transport coefficients is shown in Fig.~\ref{fig:16}(b). Their overall behavior is qualitatively similar to that observed for the parallel-plate and circular-tube geometries in the diffusion-limited regime, indicating that the same transient transport mechanisms persist in the square-rod geometry. The more complex pore structure produces a more pronounced non-monotonic response, particularly in phase--lag. We reported similar behavior previously for the advection-suppression transfer function $\Nnd$ and the spectral Sherwood number $\Shnd$ in Ref.~\cite{Hamid2023}.

At high frequency, the magnitude of $\Nx$ decreases because rapid forcing leaves insufficient time for reaction-induced concentration inhomogeneities to modify the effective advective flux. In contrast, $|\Shnd|$ increases because rapid forcing reduces the diffusion length near reactive interfaces and generates steeper concentration gradients. Its phase approaches $\pi/4$, characteristic of diffusion-controlled interfacial transport. These trends are consistent with the Warburg-type response identified analytically for the canonical geometries in Sec.~\ref{sec:Sec_04}.

The reactivity-bias transfer function $\Rx$ further illustrates the transient coupling between flow and interfacial reaction. In the low-frequency limit, it approaches a constant value that is proportional to $\Pe$, reflecting the pseudo-steady redistribution of solute by flow. As frequency increases, its magnitude decreases according to the diffusion-limited response, indicating that flow becomes progressively less effective at redistributing solute toward or away from reactive regions before the concentration field changes.

Overall, Figs.~\ref{fig:15} and \ref{fig:16} show that the transient transport mechanisms identified analytically for parallel plates and circular tubes persist in a more complex porous geometry. The framework recovers established pseudo-steady dispersion in the non-reactive limit and captures the additional frequency-dependent coupling among advection, diffusion, and interfacial reaction in the perfectly reactive limit, demonstrating its applicability beyond the canonical geometries considered in Sec.~\ref{sec:Sec_04}.

    \section{\label{sec:Sec_07}Conclusions}

In this work, we developed an omni-temporal framework that simultaneously describes advection, diffusion, and heterogeneous interfacial reaction in porous media across asymptotic and pre-asymptotic time scales. A single Robin boundary condition spans the full range of interfacial reactivity, from the non-reactive ($\Da=0$) to the perfectly reactive ($\Da\to\infty$) limit. Volume averaging of the frequency-domain pore-scale advection-diffusion equation, together with a closure ansatz relating the deviation concentration field to the reactive driving force and macroscale concentration gradient, yields a upscaled advection-dispersion-reaction equation. The resulting formulation is governed by four frequency-dependent transfer functions: the dispersion tensor $\Dnd$, the advection-suppression vector $\Nnd$, the spectral Sherwood number $\Shnd$, and the reactivity-bias vector $\Rnd$. The reactivity-bias vector, introduced here, quantifies how flow redistributes solute toward or away from reactive interfaces prior to reaction and thereby characterizes the coupling between advection and interfacial reaction that is not explicitly included in classical dispersion formulations.

The framework also provides insight into the physical origin of upscaled transport coefficients. In particular, the results show that tortuosity is not solely a geometric property but also depends on the transport process and interfacial condition. The tortuosity tensor vanishes in the perfectly reactive limit because reactive surfaces act as solute sources or sinks rather than solely as barriers to diffusion. Thus, the contribution of tortuosity to effective transport depends on the interfacial boundary condition used and its reactivity in addition to the pore geometry.

Analytical solutions of the closure problems for fully developed flow between parallel plates and through circular tubes reveal distinct kinetics-limited and diffusion-limited transport regimes. In the low-frequency limit, omni-temporal dispersion coefficient recovers the classical Taylor--Aris scaling, establishing consistency with asymptotic dispersion theory. As frequency increases, transverse diffusion has progressively less time to redistribute solute across streamlines, causing the hydrodynamic contribution to dispersion to diminish relative to molecular diffusion. The dispersion coefficient also exhibits a non-monotonic phase response, providing a characteristic signature of pre-asymptotic transport that is absent from pseudo-steady dispersion models. The remaining transfer functions exhibit distinct high-frequency responses depending on the reaction regime. In the kinetics-limited regime, $\Nnd$ and $\Rnd$ exhibit capacitive $\omega^{-1}$ scaling with a phase approaching $\pi/2$, whereas $\Shnd$ approaches a frequency-independent plateau. In the diffusion-limited regime, $\Nnd$, $\Shnd$, and $\Rnd$ exhibit characteristic Warburg-type behavior associated with transient diffusion near the reactive interface. In particular, $|\Shnd|$ grows as $\omega^{1/2}$, demonstrating that the instantaneous transient interfacial mass-transfer response can exceed its corresponding steady diffusion-limited value.

The practical significance of these frequency-dependent effects is demonstrated by comparing breakthrough curves for Gaussian solute pulses between non-reactive parallel plates with direct numerical simulations. For long-duration pulses, both the pseudo-steady and omni-temporal formulations approach the DNS results, consistent with the validity of asymptotic dispersion theory in the long-time limit. For short-duration pulses, however, the pseudo-steady formulation substantially overpredicts both propagation velocity and axial spreading. Accounting for the frequency-dependent magnitude and phase of the dispersive response substantially improves the predictions of the omni-temporal framework. The remaining discrepancies arise primarily from back diffusion associated with long-tailed concentration distributions and from entry-region effects. In particular, back diffusion generates a diffusive flux opposite to the mean flow and thereby reduces downstream solute propagation. Incorporating these non-periodic effects more directly into the upscaled description remains an important direction for future development.

The framework is further applied to a model porous medium consisting of a periodic array of square rods under cross flow. In the non-reactive limit, the calculated dispersion coefficients recover established pseudo-steady results at low frequency and transition progressively to transient behavior as the frequency increases. In the perfectly reactive limit, the framework additionally captures the frequency-dependent advection-suppression, interfacial mass-transfer, and reactivity-bias responses. Interfacial reactions also suppresses both longitudinal and transverse dispersion relative to the non-reactive limit. These results demonstrate that the principal transient transport mechanisms identified analytically for parallel plates and circular tubes persist in more complex porous geometries.

The scope of the present framework is defined by several physical assumptions. The flow is assumed to be steady, laminar, and incompressible. Although the initial volume-averaging derivation does not require a specific velocity boundary condition at the solid/liquid interface, the uniqueness analysis and subsequent results are restricted to impermeable interfaces. The porous medium is assumed rigid and macroscopically homogeneous, with an interfacial geometry that remains fixed in time, and the analysis requires clear separation among the pore, RVE, and system length scales. Species transport is assumed to be dilute and pseudo-binary, with a constant diffusion coefficient and negligible homogeneous reaction, while electromigration is neglected. Extensions to permeable or evolving interfaces, deformable pore structures, homogeneous reactions, and more general multicomponent transport provide natural directions for future work.

Overall, the omni-temporal framework provides a unified description of reactive solute transport from the asymptotic to the pre-asymptotic regime while spanning the full range of interfacial reactivity. Because pore-scale transport requires a finite time to respond to changing conditions, the macroscale flux retains a memory of the preceding driving forces. In the present formulation, this memory is encoded by the frequency-dependent transfer functions, whose corresponding time-domain memory kernels follow directly through inverse Fourier transformation without requiring a prescribed functional form. The framework therefore connects pore-scale dynamics to macroscale transient transport while recovering conventional asymptotic behavior in the long-time limit.

\section*{Acknowledgements}
The U.S. Office of Naval Research (Award no. N00014-22-1-2577) and the U.S. National Science Foundation (Award no. 1931659) supported this research.  Graduate student support from the department of Mechanical Science and Engineering at University of Illinois Urbana-Champaign (UIUC) is also acknowledged. 

\section*{Data availability}
There are no publicly available research data or software supporting this paper. Requests for further information or data should be sent to the authors.

    \bibliographystyle{apsrev4-2} 
    \bibliography{Z01_References} 

    \clearpage
    \includepdf[pages={{},-}]{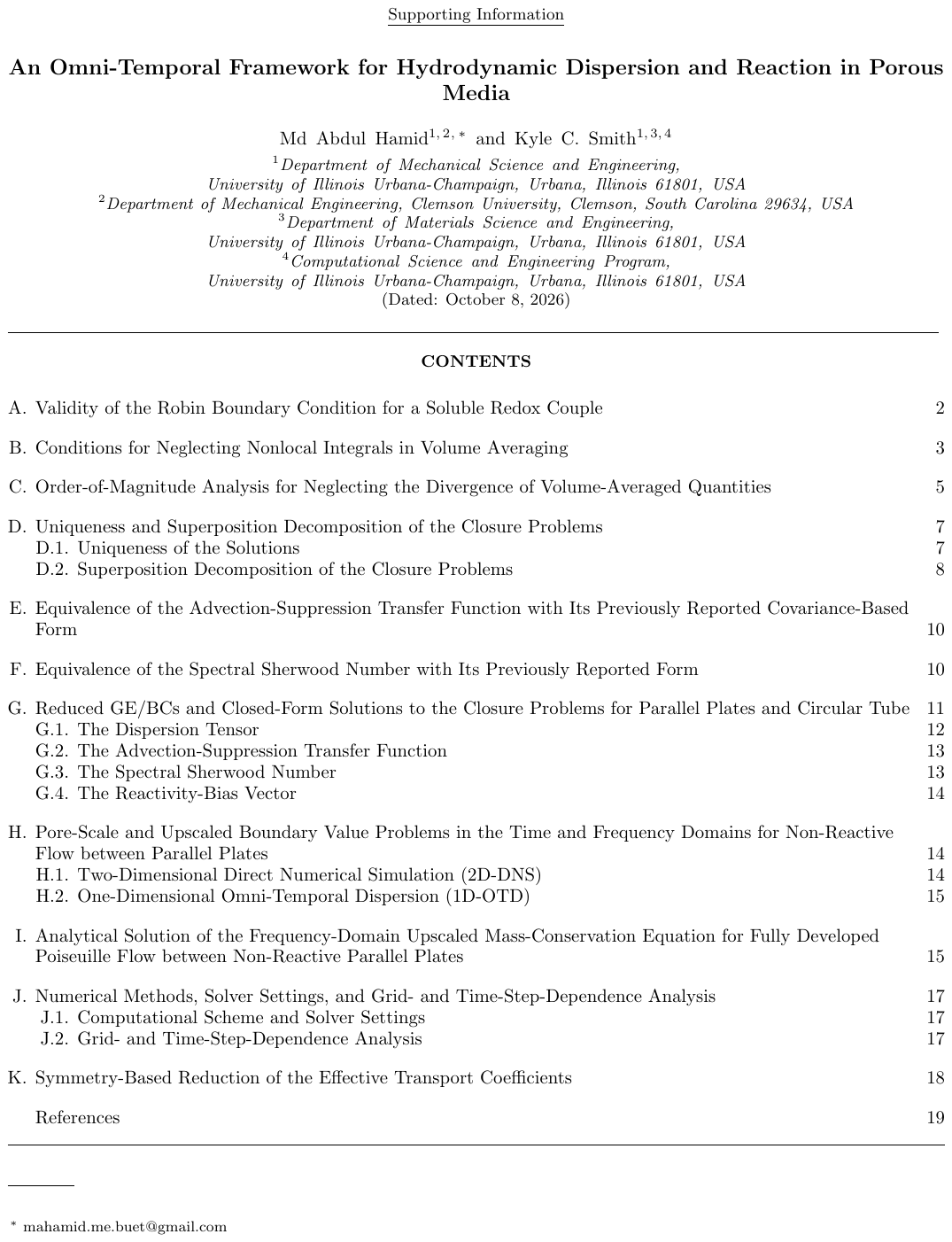}
    
\end{document}